\documentclass[iop]{emulateapj}
\usepackage{natbib}
\usepackage{graphicx}
\usepackage{subfigure}
\usepackage{amsmath}
\usepackage[utf8]{inputenc}
\usepackage{float}
\usepackage [english]{babel}
\usepackage [autostyle, english = american]{csquotes}
\MakeOuterQuote{"}

\shorttitle{MMR, Exoplanet Detection \& Characterization}
\shortauthors{Tabeshian \& Wiegert}

\begin{document}

\title{Detection and Characterization of Extrasolar Planets through Mean-Motion Resonances. I. Simulations of Hypothetical Debris Disks}

\author{Maryam Tabeshian\altaffilmark{1} and Paul A. Wiegert\altaffilmark{1,2}}
\affil{$^1$Department of Physics and Astronomy, The University of Western Ontario, London, ON, Canada, N6A 3K7 \\ 
$^2$Centre for Planetary Science and Exploration, The University of Western Ontario, London, ON, Canada, N6A 3K7 \\ \\
\textit{Published in The Astrophysical Journal (ApJ, 818, 159), February 17, 2016}}

\email{mtabeshi@uwo.ca}

\begin{abstract}
The gravitational influence of a planet on a nearby disk provides a powerful tool for detecting and studying extrasolar planetary systems. Here we demonstrate that gaps can be opened in dynamically cold debris disks at the mean-motion resonances of an orbiting planet. The gaps are opened away from the orbit of the planet itself, revealing that not all disk gaps need contain a planetary body. These gaps are large and deep enough to be detectable in resolved disk images for a wide range of reasonable disk-planet parameters, though we are not aware of any such gaps detected to date. The gap shape and size are diagnostic of the planet location, eccentricity and mass, and allow one to infer the existence of unseen planets, as well as many important parameters of both seen and unseen planets in these systems. We present expressions to allow the planetary mass and semimajor axis to be calculated from observed gap width and location.\\
\end{abstract}


\section{Introduction}
\label{Sec:Intro}

% P1
The detection of excess infrared radiation from a number of stars indicate the presence of debris disks around them, some of which, such as those around Fomalhaut and Beta Pictoris, have been observationally resolved. The first extrasolar debris disk ever discovered was found around Vega using \textit{IRAS}, from the thermal emission of circumstellar dust which revealed a strong infrared excess beyond $12 ~\mu$m \citep{Aumann84}. The excess radiation was immediately linked to the possible presence of solid dust particles with radii greater than $0.12$ cm and temperatures of approximately $85$ K, located at a mean distance of $85$ AU from the nearby main sequence star $\alpha$ Lyrae and which were believed to be debris left-overs from the formation of this stellar system \citep{Aumann84}. This discovery sparked interest in studying debris disks outside our own solar system and the first optical image of an exosolar debris disk emerged later in the same year. Using ground-based optical coronagraphy, Smith \& Terrile were able to directly image a flattened disk of cold, solid material around $\beta$ Pictoris \citep{Smith84}.

% P2
Interactions between planets and disks have been studied in great detail for gas-rich disks with the main motivation being understanding planet formation \citep[see the review paper by][]{Kley12}. The study of geometric structures in protoplanetary disks has offered insights into the formation and evolution of planetary systems. Such structures, which are mostly believed to be signposts of planet formation, reveal themselves as density variations across the disk and have been inferred from Spectral Energy Distributions and high-resolution millimeter and submillimeter interferometry observations. There is strong theoretical and observational evidence for gaps \citep[see for instance,][]{Debes13} as well as density enhancements that appear as complex features such as spiral patterns \citep[e.g.][]{Brown09, Muto12, Juhasz14} and dust traps \citep[e.g.][]{Isella13, vanderMarel13} in protoplanetary disks and that can create asymmetric structures. Observations of protoplanetary disks have revealed that asymmetric disks are common. Such asymmetries are interpreted as being either due to density perturbations of a stellar or planetary companion \citep[e.g.][]{Kraus13} or having a geometric nature \citep[e.g.][]{Brown09, Jang-Condell13}. Investigating the sources of density enhancements and depletions can result in a better understanding of the processes of formation and evolution of single and multiple planetary systems in gas-rich disks.

% P3
In a protoplanetary disk, drag against the gas causes solid particles to collapse into a dynamically cold disk in which the particle orbital eccentricities and inclinations are very low. The gas is eventually blown away by stellar radiation once the star is born and leaves behind a nearly circular, coplanar solid particle disk.This disk may include planets as well as smaller planetesimals or other bodies, like the solar system's asteroid belt. Dynamical interactions of planets with these second generation disks have not been studied as extensively as gas- rich disks. In this paper, we shall focus on this later stage in which little or no gas remains. Such gas-poor disks include for example, the Fomalhaut \citep{Cataldi15}, Vega \citep{Wilner02} and $\beta$ Pictoris \citep{Kalas95} disks. Structures in the Fomalhaut disk, for example, cannot be due to gas-dominated processes because of its low gas content \citep{Cataldi15}, but must arise from other processes.

% P4
We will show that the interaction of a planet with a debris disk can create structures that are not radially symmetric about the star. Non-axisymmetric structures have been commonly observed in debris disks. For instance, observations of the debris disk around $\beta$ Pictoris revealed a warp in the inner disk around $\sim 70-150$ AU \citep[e.g.][]{Burrows95, Heap00}. Dynamical modeling had suggested that the warp could be explained by a misaligned planet \citep[e.g.][]{Mouillet97, Augereau01} which was later confirmed when a $9\pm3$ Jupiter-mass planet was found on an inclined orbit $8-9$ AU from the central star \citep{Lagrange10}. Asymmetries in debris disks have also been attributed to mean-motion resonances (MMRs). For instance, \textit{N}-body simulations of a collisional debris disk by \cite{Nesvold15} show a peak in the disk's surface brightness at 1:1 MMR with the planet. A second peak is observed in their simulation of a 3 Jupiter-mass planet and falls between the 3:2 and 2:1 MMR with the planet, indicating a depletion of planetesimals at the two resonances. Formation of overdensities in debris disks has also been explained by migrating dust \citep[e.g.][]{Wilner02} or migrating planets \citep[e.g.][]{Wyatt03}. While migrating inward due to Poynting–Robertson (PR) drag, dust grains can become trapped in resonances with a planet interior to their orbit or get scattered, thus forming asymmetric structures in the disk. The same scenario applies to particles that are captured in a migrating planet's resonances.

% P5
Structures in dust disks in which particles are strongly affected by the radiation pressure of the central star as well as the PR drag have been investigated by some authors \citep[see for instance][]{Kuchner03, Wyatt06, Krivov07}. However, not much emphasis has been placed on the dynamical interactions of planets with planetesimal belts with regards to understanding of the MMR gaps and how they can be used to extract information about the planets causing them. For instance,  \cite{Chiang09} briefly discuss gap formation at MMRs with a planet interior to the Fomalhaut disk in an attempt to constrain the mass of Fom b, but do not provide further details on how such gaps could yield measurements of planetary parameters. On the other hand, \textit{N}-body numerical simulations by \cite{Reche08} show gap structures similar to what we shall discuss in the present work, but were not addressed by the authors. Here we address the question of what structure might be induced in a gas-poor extrasolar planetesimal disk by a non-migrating planet, and in particular what observational signatures might indicate the presence of an unseen planet in the system and how they can be used to constrain the planet's mass and orbital parameters. It must be noted that for the rest of this paper, whenever the term “debris disk” is used, it means disks with particles having size distributions in the range of about $1$ m to $100$ km that are nearly unaffected by the central star's radiation pressure. Such particles can be gravitationally perturbed due to the presence of one or more planets in the system which could result in either their removal (like the Kirkwood Gaps in our solar system) or accumulations (like the Hilda family of main belt asteroids). Studying these resonances is not only an indirect way of detecting unseen extrasolar planets, it can also help put constraints on some parameters of the planets which are creating them, such as their mass, semimajor axis and eccentricity.

% P6
The astonishingly detailed image of a disk around HL Tau, a Sun-like star approximately 450 light-years away in the constellation Taurus that was recently obtained by the Atacama Large Millimeter Array (ALMA), is a perfect example of structures being formed in disks due to planets \citep{NRAO14}. Although the disk imaged around HL Tau is a protoplanetary disk different from the second generation debris disks we concentrate on in the present work, the ALMA image of HL Tau illustrates the increasing resolving power that can be achieved with state of the art telescopes. With ALMA soon starting its full operation, more detailed images of debris disks will become available, revealing more and more detailed structures such as those discussed here. 

% P7
This paper is organized as follows: In Section \ref{Sec:MMR}, we briefly describe the dynamics of MMRs and the theoretical calculations of their maximum libration widths. Our method is presented in Section \ref{Sec:Sims} in which we explain the code that is used to generate the simulations as well as the initial conditions. We present the different results obtained for interior versus exterior MMRs in Section \ref{Sec:Results}, followed by discussions in Section \ref{Sec:Discussions}. Finally, a summary and conclusions are provided in Section \ref{Sec:SummConc}.

\section{The Dynamics of Mean-Motion Resonances}
\label{Sec:MMR}

% P1
The existence of gaps in the solar system's asteroid belt was first noted by American astronomer Daniel Kirkwood in 1867 who saw non-uniformities in the number distribution of asteroids in the main belt as a function of semimajor axis, with some ranges having few or no asteroids \citep{Kirkwood67}. We now understand that gravitational interactions between the asteroids and Jupiter result in the removal of planetesimals from these orbits, making gaps to appear in the disk where a particle's orbital period would be a simple fraction of that of Jupiter's \citep{Murray99}. The close link between orbital period of an asteroid and its semimajor axis means that these "MMRs" occur over narrow ranges of semimajor axes, often depleting them of their original complement of bodies.

\begin{figure*}
    \centering
    \includegraphics[totalheight=0.6\textheight]{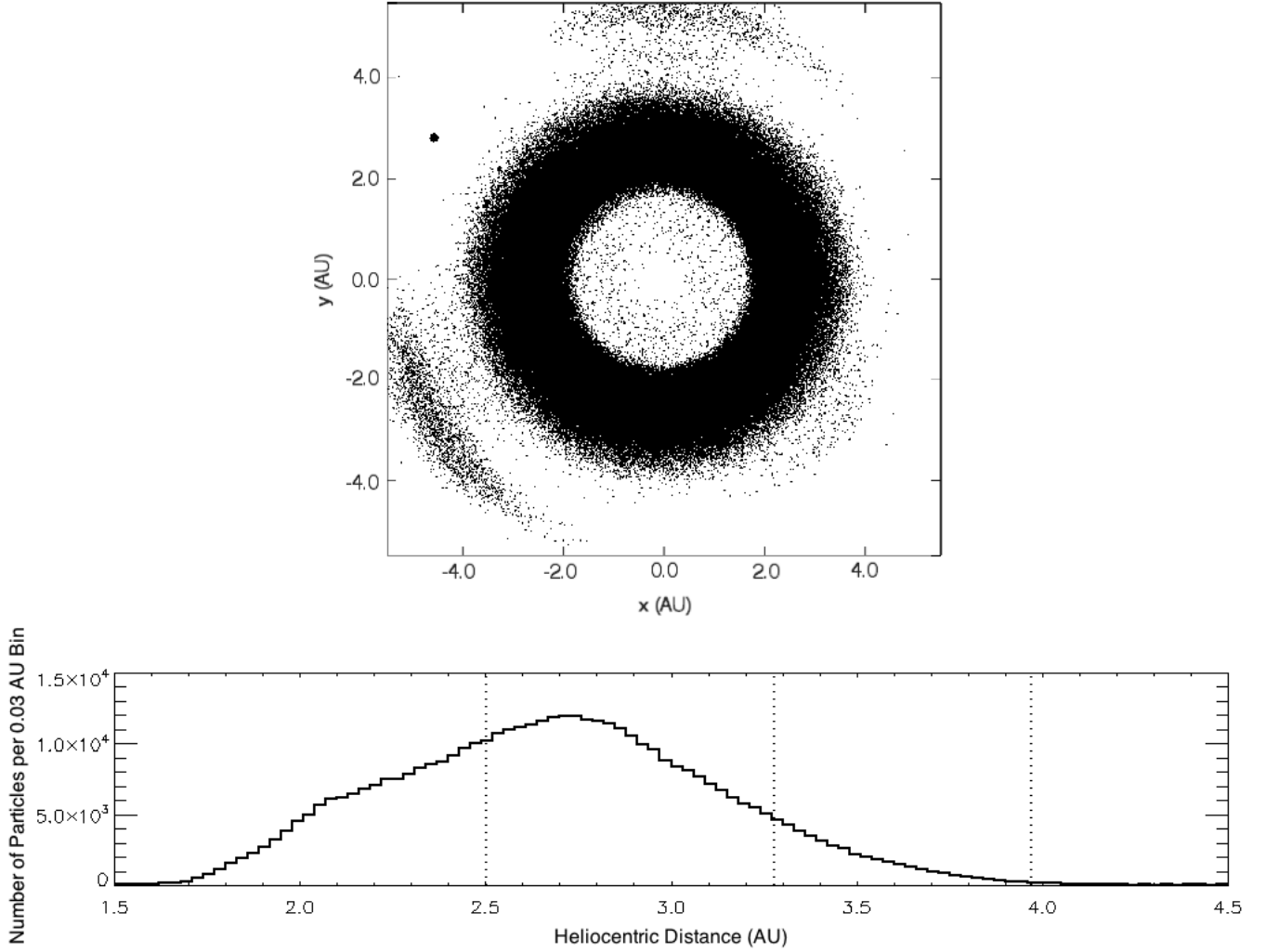}
    \caption{Orbits of known solar system asteroids interior to Jupiter's as of 2015 June 27 plotted using data obtained from the Asteroids Dynamic Site \citep{AstDyS}. No Kirkwood gap due to MMR with Jupiter (shown with the black circle) can be seen in the disk (top panel) or the histogram (bottom panel) which shows the distribution of asteroids per distance from the Sun. This is due to the eccentricities of the asteroids that bring them in and out of the gaps. The theoretical locations of the three strongest resonances are shown on the histogram with vertical dotted lines.}
    \label{Fig:MAB}
\end{figure*}

% P2
However, if one were to take an image of our asteroid belt from outside the solar system, the Kirkwood gaps would not be observable due to the eccentricity of the asteroids blurring the edges of the gaps (see Figure \ref{Fig:MAB}). Though asteroids are removed from a number of resonances in the main belt, the eccentricities of the remaining asteroids are large enough to blur the edges of the resonances and make the gaps invisible. We will show that this is not always the case, and that under realistic conditions, visible gaps can be opened in particle disks by resonances. This means that gaps in extrasolar disks do not necessarily contain planets, as often assumed. Planets can also generate other resonant structures that both indicate their presence and provide diagnostic information about their mass, eccentricity and position. Therefore, not only is the observation of gaps in debris disks an indirect way of detecting undiscovered extrasolar planets, it can also be used to constrain some parameters of the perturbing body if the gap widths and locations can be measured.

%% Figure 1

% P3
Two objects are said to be in MMR if the following relation holds:

\begin{equation}
\label{Eq:MMR_Def_n}
\frac{n}{n^\prime} = \frac{p+q}{p} ~,
\end{equation}

\noindent where $n$ and $n^\prime$ are their mean-motions ($=\frac{2\pi}{T}$, with $T$ the orbital period), and $p$ and $q$ are positive integers, with $q$ denoting the order of resonance. The primed and unprimed quantities are the orbital elements of the particle being perturbed (the "asteroid") and the perturbing body ("the planet"), respectively. Equation \ref{Eq:MMR_Def_n} can, equivalently, be written in terms of the two objects' semimajor axes, $a$ and $a^\prime$, as:

\begin{equation}
\label{Eq:MMR_Def_a}
a^\prime = \left(\frac{p+q}{p}\right)^{\frac{2}{3}k} ~ a ~,
\end{equation}

\noindent where $k=+1$ for exterior resonance (i.e. $a^\prime > a$) and $k=-1$ for interior resonance (i.e. $a^\prime < a$). 

% P5
Following the discussion and derivations presented in \cite{Murray99}, to lowest order the resonant argument of the disturbing function, $\phi$, can be written as:

\begin{equation}
\label{Eq:phi}
\phi = j_1 \lambda^\prime + j_2 \lambda + j_3 \varpi^\prime + j_4 \omega ~,
\end{equation}

\noindent where $\lambda$ and $\lambda^\prime$ are the mean longitudes and $\omega$ and $\varpi^\prime$ are the arguments of periapse. Also, $j_1=p+q$, $j_2=-p$ and $j_3$ and $j_4$ are either zero or $-q$, depending on the relative locations of the two objects. 

\begin{figure*}
    \centering
    \includegraphics[totalheight=0.4\textheight]{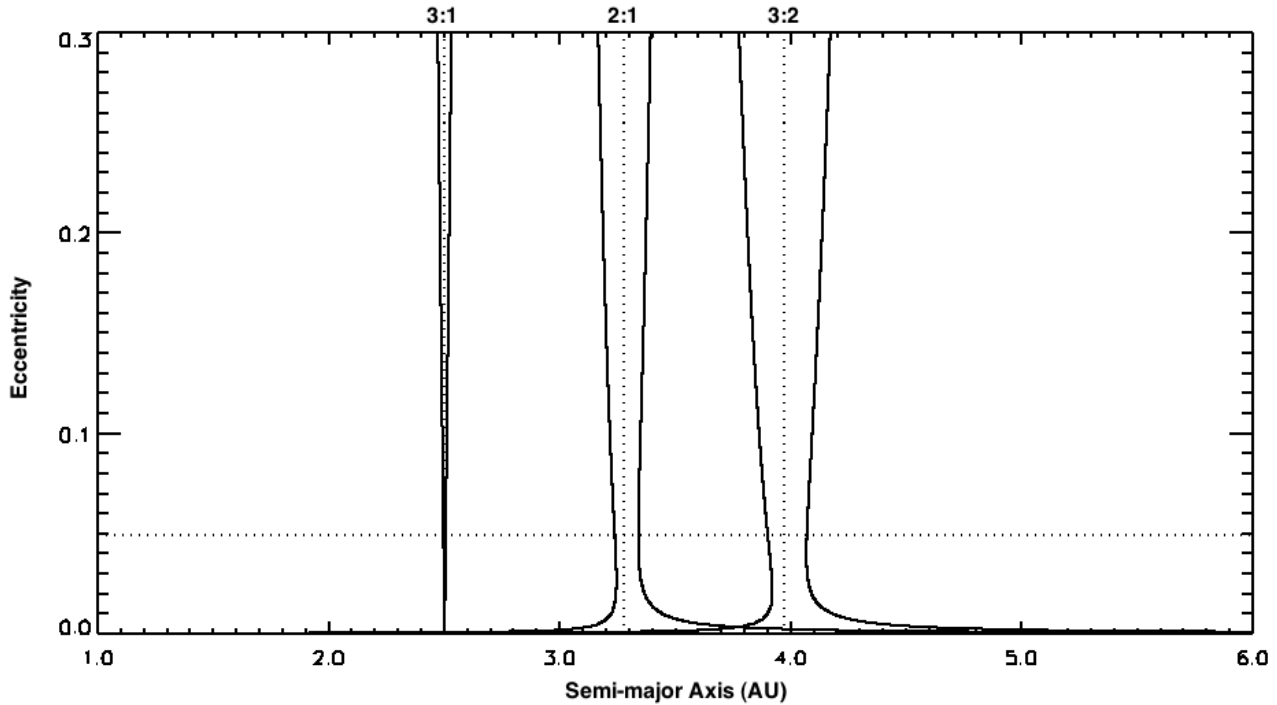}
    \caption{Maximum libration widths for 2:1, 3:2, and 3:1 main belt resonances with Jupiter. The vertical dotted lines are the locations of nominal resonances (calculated using Equation \ref{Eq:MMR_Def_a}) and the horizontal dotted line shows Jupiter's eccentricity.}
    \label{Fig:AnalyticPlot}
\end{figure*}

% P6
When a particle in a debris disk orbits in a MMR with a perturbing body, such as a planet, its orbit is perturbed in a consistent manner when the relative planet-asteroid geometry repeats itself. This often destabilizes the smaller body so that it either collides with the planet, crashes into the star, goes into a highly elliptical orbit or gets ejected. In either case, a gap forms in the disk whose width, $\delta a^\prime_{max}$, can be approximated by the maximum libration width of its resonance. This can be calculated analytically at low eccentricities using Equation \ref{Eq:delta_ap}:

\begin{eqnarray}
\frac{\delta a^{\prime}_{max}}{a^{\prime}}  &=& \pm \left\{\left(\frac{16|C^{\prime}_r|}{3n^{\prime}}e^{\prime} \right)^{\frac{1}{2}} \left(1+\frac{1}{27j^2_2e^{\prime^3}} \times \frac{|C^{\prime}_r|}{n^{\prime}}\right)^{\frac{1}{2}}\right\} \nonumber \\
&& -\frac{2}{9j_2e^{\prime}} \times \frac{|C^{\prime}_r|}{n^{\prime}} ~,
\label{Eq:delta_ap}
\end{eqnarray}

\noindent where:

\begin{equation}
\label{Eq:C'_r}
C^\prime_r = \left(\frac{GM}{a^{\prime^2} ~ a ~ n^\prime}\right) \times f_d (\alpha) ~,
\end{equation}

\noindent with $G$ denoting the Universal Gravitational Constant, $M$ the mass of the perturbing body, $f_d (\alpha)$ a term containing Laplace coefficients and coming from three-body expansion of the two objects' orbital elements and $\alpha = (\frac{a}{a^\prime})^{k}$. Equation \ref{Eq:delta_ap} is given as Eq. (8.76) in \cite{Murray99}. Figure \ref{Fig:AnalyticPlot} shows the maximum libration widths at various eccentricities for the 2:1, 3:2, and 3:1 interior resonances with Jupiter. It is clear from this figure that in the case of first order resonances, the libration widths increase with eccentricity, except at very low eccentricities for which they become very large. 

% P7
A MMR affects a narrow range of semimajor axes. The resulting resonant gaps will be visible in an optical or other telescopic image of the disk if (1) the resonance tends to destabilize particles (often but not always the case), (2) the eccentricity of the particles near the edges of the resonance is small enough to keep them from blurring the edges of the gap, and (3) the width of the gap is not smaller than the resolution of the telescope. The radial excursion, $\Delta r$, of the particles in the disk, given by $\Delta r \sim 2ae$, should be smaller than the resonance width.  Thus the formation of MMR gaps is favored in dynamically cold particle disks, which have small eccentricities. It is the relatively high average eccentricities of asteroids that obscure the Kirkwood gaps in our own main asteroid belt. From Figure \ref{Fig:AnalyticPlot}, it is also clear that very low eccentricity disks may have very large resonant gaps since the resonance width increases sharply as the disk average eccentricity, $e^\prime_{ave}$, approaches zero; hence very dynamically cold disks are good candidates for observing particularly large resonant gaps.

\section{Simulations}
\label{Sec:Sims}

\subsection{The Method}
\label{Sec:Method}

% P1
Our simulations are performed with a symplectic integrator based on the Wisdom–Holman algorithm \citep{Wisdom91}. A fixed timestep of 50 days is used for all simulations. The output is recorded at 10,000-year intervals and the total simulation time is taken to be 1 million years in length unless otherwise noted. A single planet on a circular or slightly elliptical orbit perturbs the disk, which orbits a 1 solar-mass star. Particles are removed if they have a close encounter with the planet or reach a distance less than 10 solar radii or greater than 1000 AU.

% P2
The gravitational effects of the star and planet are included but interactions between the particles themselves are ignored. Thus our simulations are applicable to low-mass debris disks, where the mass of the disk is much less than that of the planet. We also neglect the effects of the PR drag for simplicity; hence our simulations represent gas-poor planetesimal or debris disks composed of solid bodies 1 m to 100 km across and which are relatively free of dust. The presence of strongly reflecting or emitting dust can markedly affect the appearance of a disk, particularly if collision among the dust particles is considered. For instance, a study conducted by \cite{Stark09} shows that the ring structures created by the trapping of dust in resonance with a planet are smeared out by collisional interactions of the dust particles. On the other hand, \cite{Wyatt05} has argued that the collisional lifetime of dust in debris disk candidates is short enough that dust does not drift very far before its destruction. Here we model dynamically cold planetesimal disks which are dust-poor. However, we note that the observational characteristics of such disks will be dominated by much-smaller dust (which has different dynamics) in dust-rich systems. The mechanisms of dust production and removal in these disks are complex and their modeling is outside the scope of this paper. Nonetheless, it is reasonable to expect that at least some dust-poor systems exist, and it is to those that we turn our attention here.

% P3
The simulations are performed at scales appropriate to our solar system (i.e., the perturbing planet is placed at Jupiter's semimajor axis, $\sim$ 5.204 AU from the star). The physics involved scale with distance, however, and so our results are applicable to disks and planets in general, even if located at different distances from their parent stars. For clarity then, our figures are scaled so that the planet is at a unit distance. The exception to the scalability of our results is only the timescales for opening up the gaps, which are expected to be longer for larger systems or ones with less massive central stars.

\subsection{Simulated Debris Disks}
\label{Sec:SimDD}

% P1
In order to investigate structures in debris disks that are caused by MMRs with a planet, we perform simulations of test particles in a flat disk (particle inclinations $i^\prime=0.0\degr$), containing 10,000 particles per 1 AU of the disk radial thickness. Running on a single CPU, the simulations take 12 hrs to complete for a disk with 20,000 particles and could last up to 4 days for three times the number of particles. The total simulation time also depends on the planet's mass since more particles are ejected at the beginning of the simulations with more massive planets. The initial particle eccentricities are those of the forced eccentricity induced by the planet at each particle's semimajor axis with their apses aligned with the planet's. The forced eccentricity represents the eccentricity that the particle orbits are subject to due to the simple presence of the perturbing planet. By setting the initial conditions of the disk to this value, we create a disk which will be minimally perturbed by the planet. This choice represents a scenario where the planet and the disk have been in the same relative geometry for a significant fraction of the age of the system. Other choices result in a more heavily perturbed disk, with results that are highly dependent on the choice of the initial conditions. Such scenarios might be appropriate to cases of recent planetary migration but are not considered here. We note that for the case of a planet on a perfectly circular orbit, the forced eccentricity is also zero and the disk particles are started on circular orbits.

% P2
For simplicity, we assume that the planetary system contains one planet only and we choose the planet's semimajor axis to be that of Jupiter ($a=a_J=5.204 ~AU$), though as noted earlier the choice of this scale is arbitrary and the results apply equally to disk-planet systems of all sizes. Furthermore, in all of our simulations, we assume that the planet orbits in the same plane as the disk (i.e. $i=0.0\degr$). However, we also tried some simulations with a small planet inclination ($i=i_J=1.304\degr$, Jupiter's inclination). The results show no noticeable difference between the two cases. Therefore, even if the planet does not orbit exactly in the disk plane, we still expect the same features and to obtain similar results for small orbital inclinations.

% P3
The simulations are performed for the case of both interior and exterior resonances; and in both cases, the disk is placed 1 AU away from the planet ($2.204<a_{Disk}<4.204$ for the interior resonance and $6.204<a_{Disk}<12.204$ for the exterior case). The values for the disks' inner and outer edges are chosen such that the three resonances being considered, the 2:1, 3:2, and 3:1, fall within the debris disks. A range of planet masses is used, going from as small as $1.0~ M_\oplus$ to $9.0~M_J$, where $M_\oplus$ and $M_J$ are the mass of the Earth and Jupiter respectively. Furthermore, to study the effect of the planet's eccentricity on MMR gaps, two different planet eccentricity values are considered: $e=0.0$ and $e=0.0489$, the eccentricity of Jupiter ($e_J$), though we leave further investigations of eccentricity effects to a follow-up paper.

\section{Results}
\label{Sec:Results}

% Moved from the Methods Section
As noted before, we choose our total simulation time to be a million years, which we find to be sufficient for the disks to achieve a quasi-steady state. Yet we observe gaps forming on much shorter timescales for more massive planets (e.g., only 100,000 years, for a 5 Jupiter-mass planet exterior to the disk). More massive planets open gaps more quickly as would be expected, but we have not investigated this trend in the present work. At the end of the simulations, the disks are examined for structures, particularly those that would be observationally discernible in a telescopic image. Moreover, in order to be able to compare the widths of the gaps that are produced in the simulations with the analytical calculations of the maximum libration widths through Equation \ref{Eq:delta_ap}, we make histogram plots of the number of particles per heliocentric distance. The disk is divided azimuthally into four equal segments since the MMR gaps turn out to be azimuthally asymmetric, particularly in the case of the perturbing body having zero eccentricity.

It must be noted that although we calculate what the resonance widths should be analytically for the 3:2 and the 3:1 resonances as well as the 2:1, due to the narrower gaps these resonances produce we only measure the widths of the 2:1 gaps in our simulations. However, as we shall discuss in Section \ref{Sec:Lindblad}, we do observe a narrow feature at the 3:1 resonance.

\subsection{Interior Resonance}
\label{Sec:IntRes}

\begin{figure*}
    \centering
    \subfigure[Interior resonances result in formation of two arcs of gaps in the disk. The dashed line shows the periastron of the planet's orbit and the symbols represent the three resonances considered while the filled circle is the planet. Different colors represent the different segments for which we make histograms of the distributions of particles in the disk (see Figure \ref{Fig:outerEI0-hists}).]{%
    \includegraphics[totalheight=0.25\textheight]{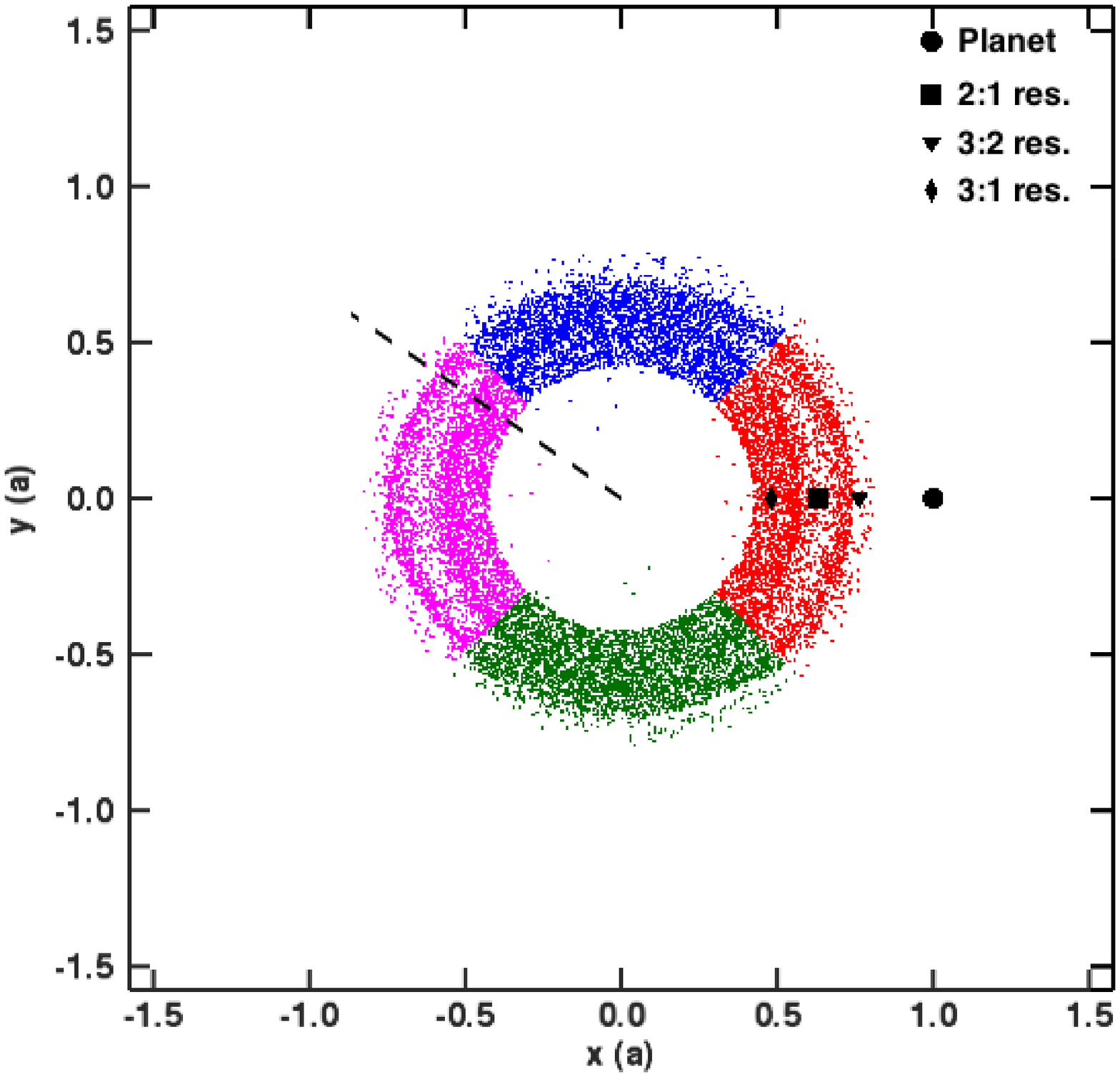}
    \label{Fig:outerEI0-disk}}
\hfill
    \subfigure[Distribution of particles in the disk and MMR structures due to a planet exterior to the disk. The colors in the first four panels correspond to the same colors in Figure \ref{Fig:outerEI0-disk} for the different segments (from top to bottom: regions separated by intervals of $\pi/2$, starting with $-\pi/4<\theta<\pi/4$). The last panel shows the overall number distribution of the particles in the disk. The dotted lines are the theoretical locations of the 2:1, 3:2, and 3:1 resonances with the dashed lines defining the width of each gap calculated analytically. A Gaussian fit is made to the top histogram where the gap is to obtain a measure of the gap width from the simulations. The bin size is $ 0.006 ~a $, in the unit of the planet's orbital radius.]{%
    \includegraphics[totalheight=0.46\textheight]{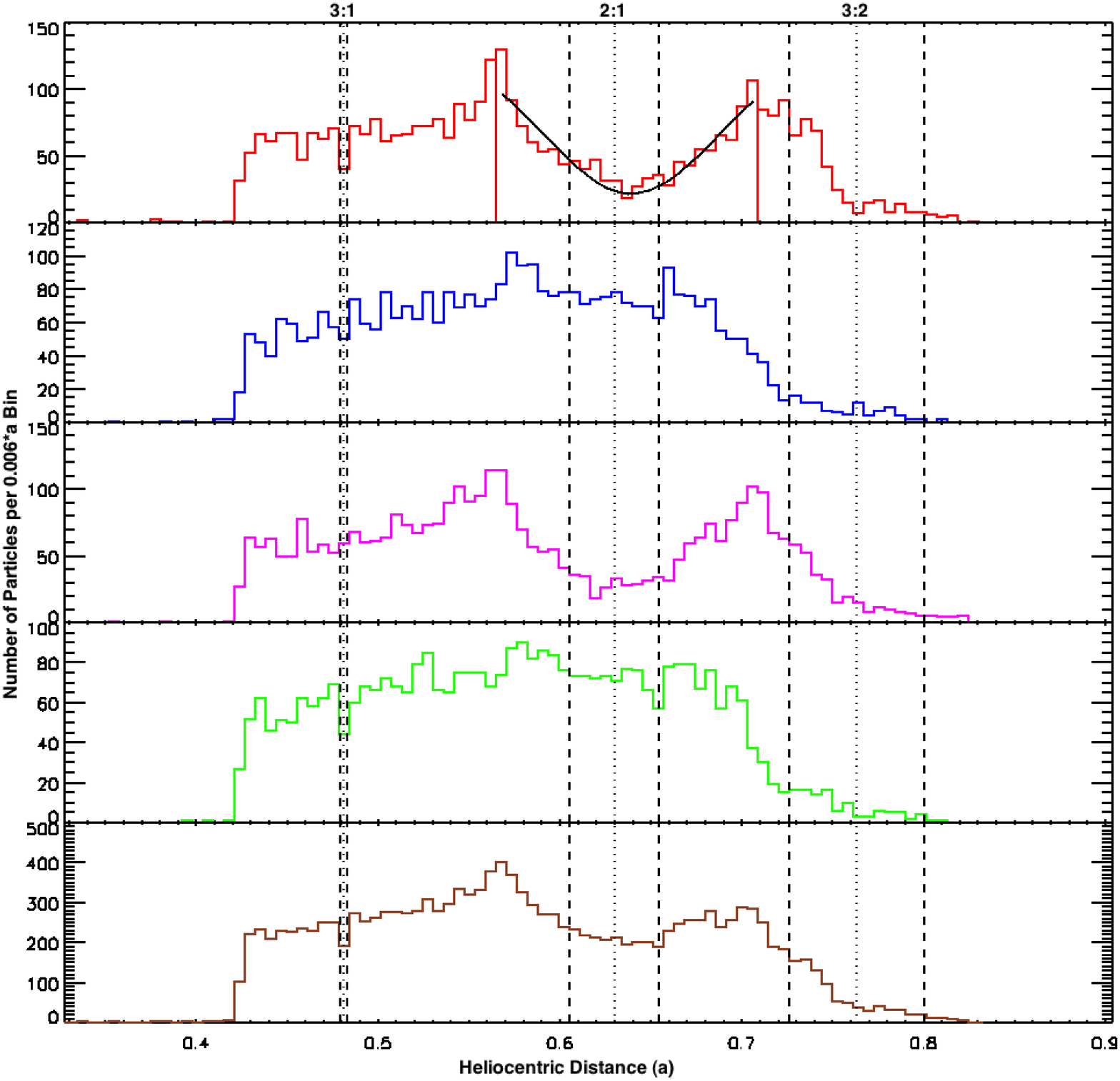}
    \label{Fig:outerEI0-hists}}
\caption{Simulation showing interior resonance structures formed by interactions of planetesimals in the disk with a $4.0~M_J$ planet on a circular orbit exterior to the disk.}
\label{Fig:outerEI0-disk-hists}
\end{figure*}

% P1
Interior resonance refers to the case in which the planet is exterior to the disk and hence the resonances occur interior to the planet's orbit. In this case, the 2:1 MMR opens gaps in two regions: one at inferior conjunction with the planet and one at opposition. This can be seen in Figure \ref{Fig:outerEI0-disk} which shows the case of a $4.0 ~M_J$ planet on a circular orbit gravitationally interacting with the disk. The locations of the three prominent resonances are marked; the gap occurs at the 2:1 MMR with the planet. If we were to follow the double-arced gap over time, we would see it move around the star at the same rate as the planet. The appearance of such a gap in an image of a disk would both indicate the presence of a planet as well as its location (that is, along a line drawn through the deepest parts of each arc), though which side of the disk the external perturber is on cannot be unambiguously determined. This double-arc shape is a result of the removal of particles from the 2:1 interior resonance \citep[see for example Fig. 8.4(a) of][]{Murray99}.  

% P2
In order to measure the width of the gap seen in the simulation, we make a histogram for each of the four colored segments (separated by intervals of $\pi/2$, starting with $-\pi/4<\theta<\pi/4$) shown in Figure \ref{Fig:outerEI0-disk} for the number of particles as a function of their distance from the central star. This is shown in Figure \ref{Fig:outerEI0-hists} where the colors of the first four histograms correspond to the same colors used in Figure \ref{Fig:outerEI0-disk} while the last histogram shows the distribution of all the particles from the four segments put together. For simplicity, we will concentrate on the gap in the region closest to the planet (marked in red in Figure \ref{Fig:outerEI0-disk} and the top histogram of Figure \ref{Fig:outerEI0-hists}).

% P3
The dotted lines in Figure \ref{Fig:outerEI0-hists} are the theoretical locations of the three resonances that are being considered, calculated using Equation \ref{Eq:MMR_Def_a}, while the dashed lines show the theoretical width of each resonant gap, obtained from Equation \ref{Eq:delta_ap}. In addition to the gaps that appear at the 2:1 MMR, the histograms also show a slight decrease in the number of asteroids in the region corresponding to the 3:1 MMR with the planet, although it is considerably narrower compared to the 2:1 resonance. The decrease at the 3:1 resonance in the sector nearest the planet is a difference of about 50\% which is 5 times the Poisson error for the bin and hence is statistically significant at this resolution. On the other hand, the 3:2 resonance which is closer to the planet cannot be seen since large numbers of particles in the outer edge of the disk have been removed due to strong gravitational interactions with the planet. To obtain a measure of the width of the 2:1 gap, we make a Gaussian fit to the histogram where the gap appears for comparison to the analytic results which we discuss in Section \ref{Sec:WvsMP}.

\begin{figure*}
    \centering
    \subfigure[Same as Figure \ref{Fig:outerEI0-disk} except that the planet's orbital eccentricity is increased to $\sim 0.05$. The extra feature at the 3:1 MMR is likely launched by Lindblad resonances and appears whenever the planet is given non-zero eccentricity.]{%
    \includegraphics[totalheight=0.25\textheight]{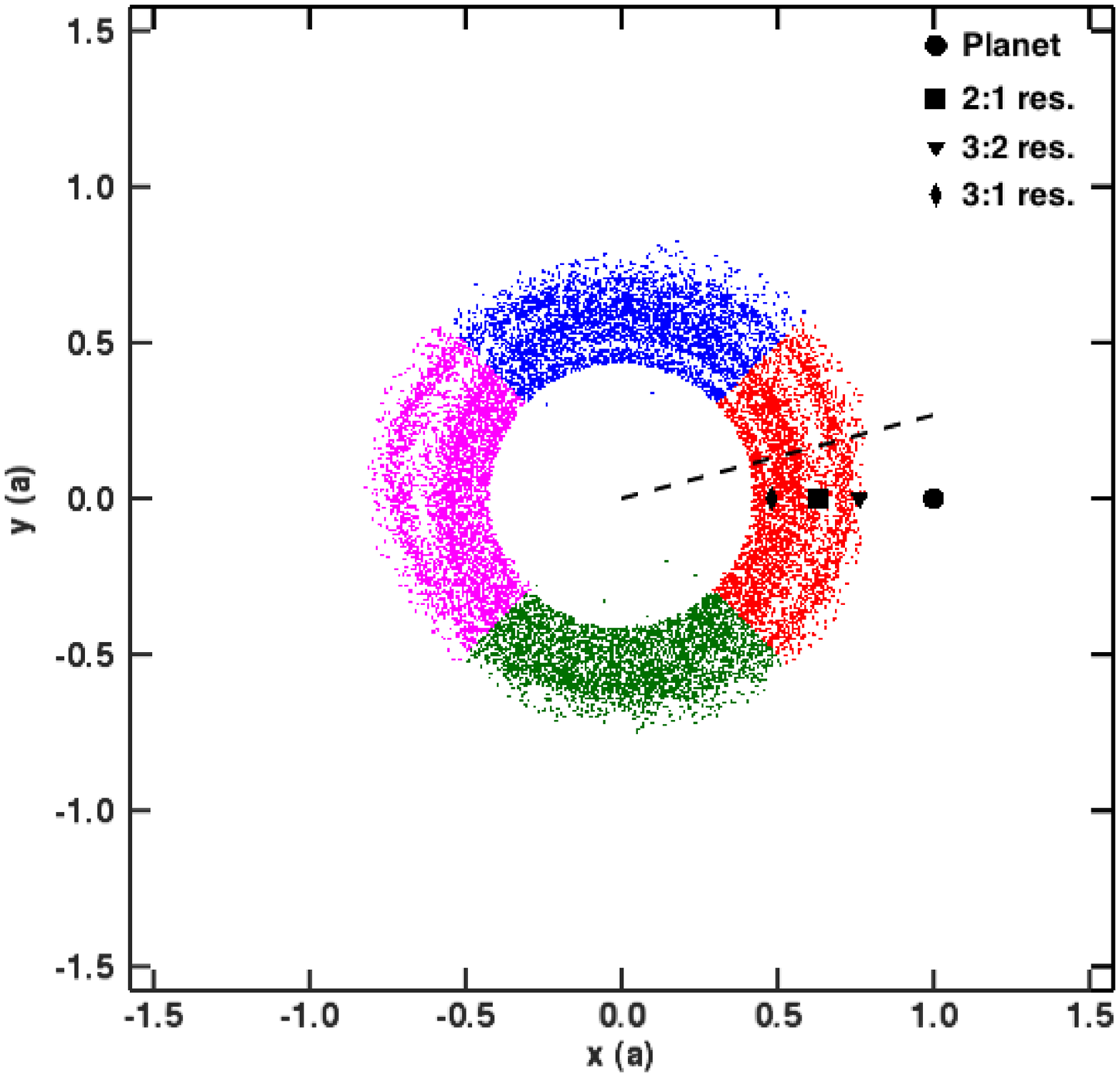}
    \label{Fig:outerERI0-disk}}
\hfill
    \subfigure[Histogram plots for the disk shown in Figure \ref{Fig:outerERI0-disk} also indicate the presence of an extra gap at the 3:1 resonance with the planet.]{%
    \includegraphics[totalheight=0.46\textheight]{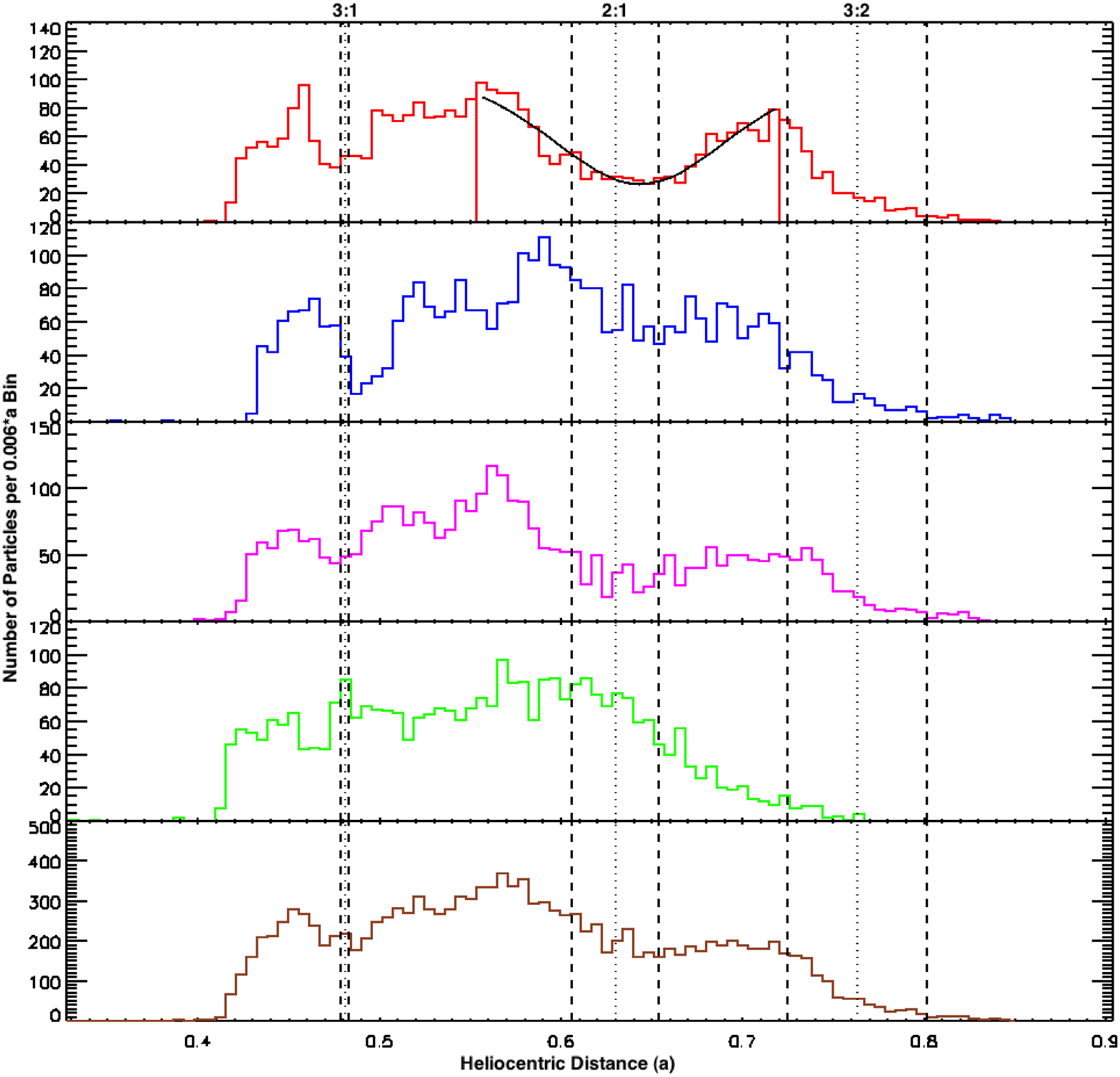}
    \label{Fig:outerERI0-hists}}
\caption{Same as Figure \ref{Fig:outerEI0-disk-hists} but with planet's eccentricity increased to $e = e_J = 0.0489$.}
\label{Fig:outerERI0-disk-hists}
\end{figure*}

% P4
Figure \ref{Fig:outerEI0-disk-hists} belongs to the case in which the planet perturbing the disk has zero eccentricity. When the planet's eccentricity is increased to 0.0489, features are seen in the disk that are similar to those of the zero eccentricity case. Additionally, the feature at the 3:1 MMR broadens. This feature, unlike the 2:1 gap, does not orbit the central star at the same rate as the planet. This is shown in Figure \ref{Fig:outerERI0-disk-hists} for a planet with $e=e_J=0.0489$. We attribute this to the generation of tightly wound spiral waves at a Lindblad resonance in the disk, a phenomenon we explore in more detail in Section \ref{Sec:Lindblad}.

\subsection{Exterior Resonance}
\label{Sec:ExtRes}

\begin{figure*}
    \centering
    \subfigure[Exterior resonance with a planet on a circular orbit results in the formation of a gap that appears as a single arc in the region of the disk closest to the planet.]{%
    \includegraphics[totalheight=0.25\textheight]{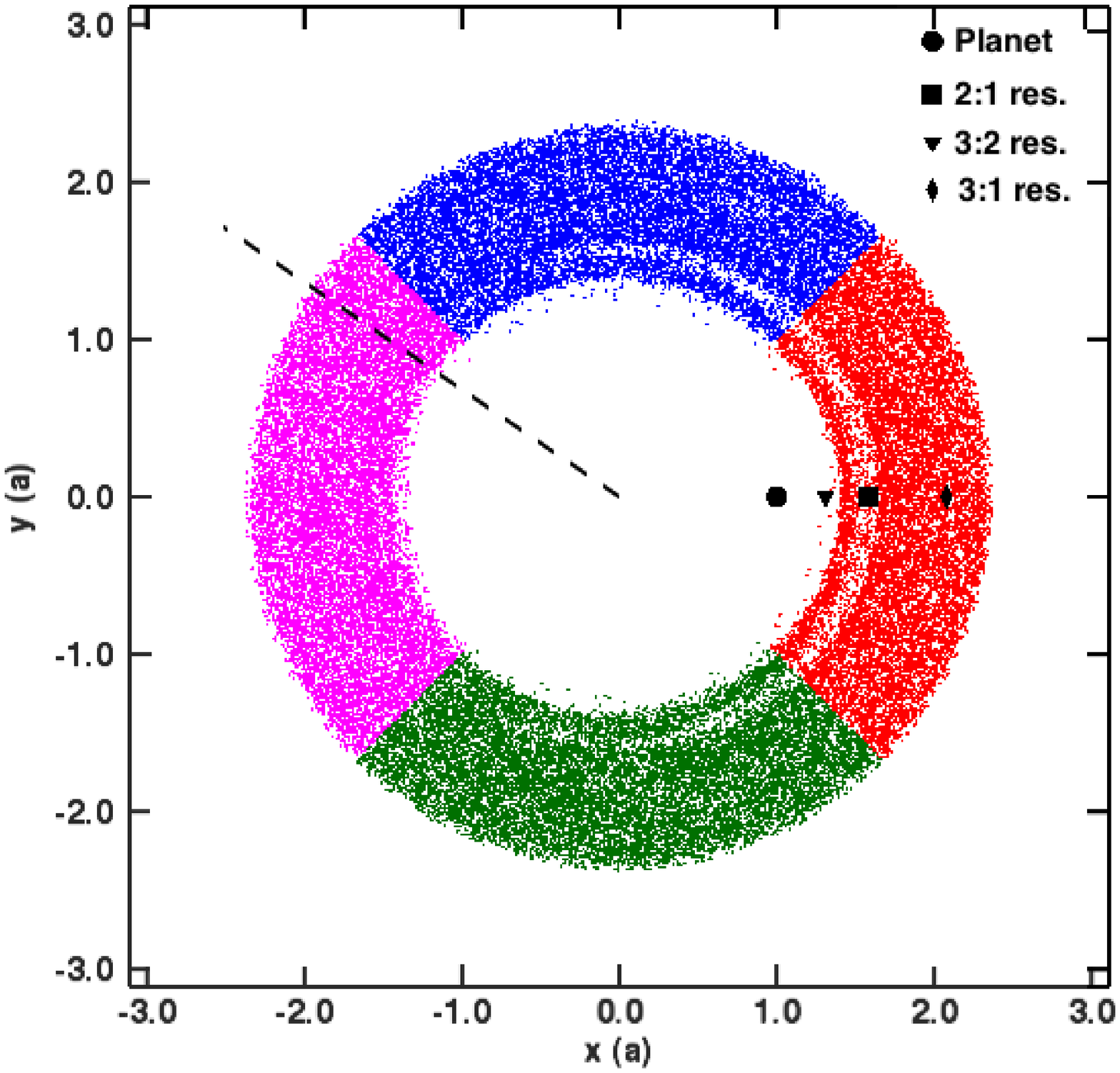}
    \label{Fig:innerEI0-disk}}
\hfill
    \subfigure[Distribution of particles in the disk for each segment marked in Figure \ref{Fig:innerEI0-disk}.]{%
    \includegraphics[totalheight=0.46\textheight]{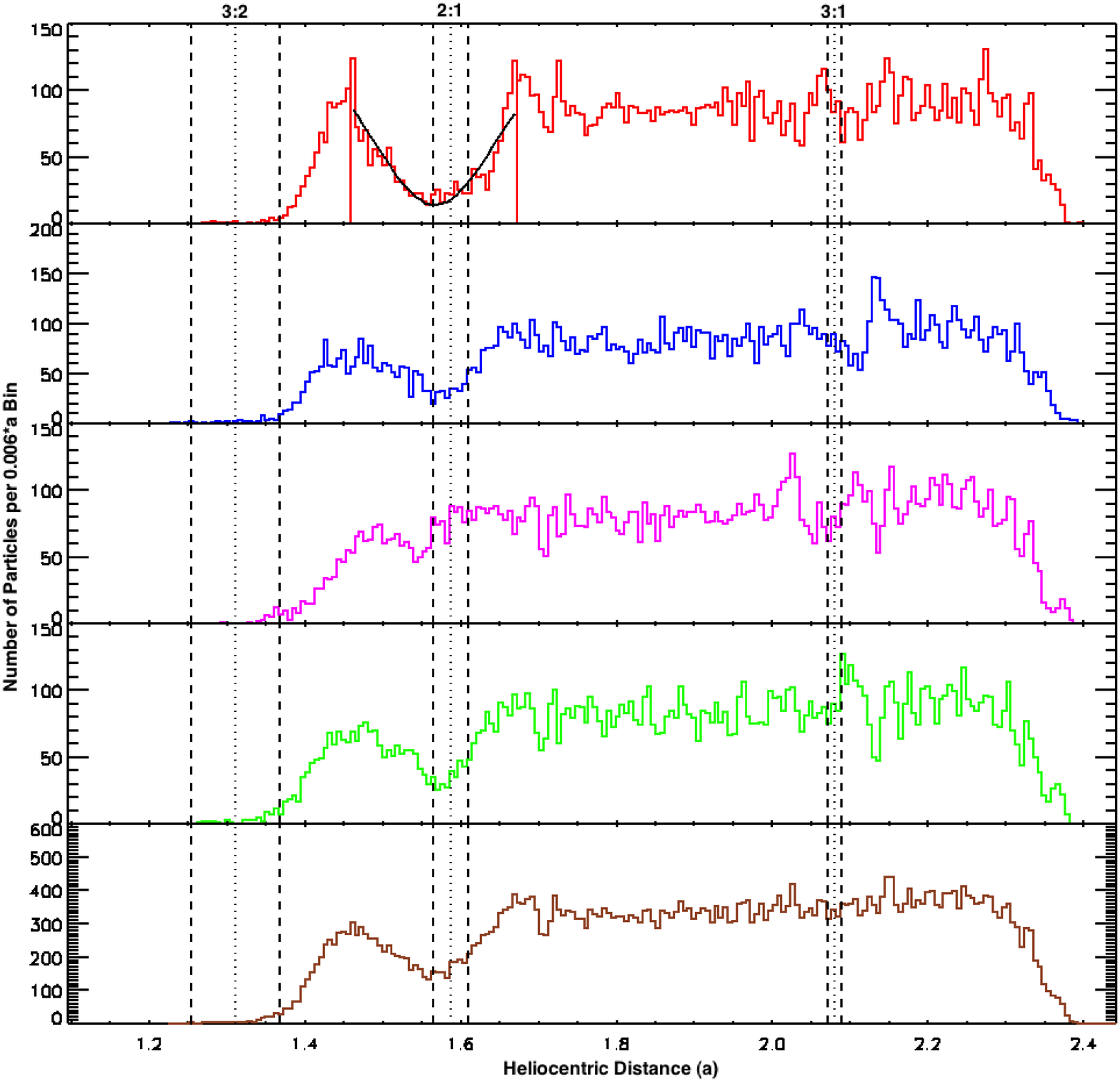}
    \label{Fig:innerEI0-hists}}
\caption{Structures formed by resonance interactions of planetesimals with a $4.0~M_J$ planet on a circular orbit interior to the disk.}
\label{Fig:innerEI0-disk-hists}
\end{figure*}

% P1
Resonance structures are different depending on whether the disk is internal or external to the orbit of the planet. For the case of exterior resonance (i.e., the planet being interior to the disk), we find the shape of the gap to be different. Instead of forming two arc-like sectors, the gap formed due to resonant interactions with a planet on a circular orbit is a single arc whose center aligns with the planet. This is shown in Figure \ref{Fig:innerEI0-disk} for a planet with $M=4.0 ~M_J$. The arc again co-rotates with the planet and provides an indicator of the location of the perturbing body even if it were not visible in an image of the disk. We note again that the most prominent gap appears where the 2:1 MMR with the planet is (shown by the filled square). The different shape of the gap can also be explained in the same manner as for the interior resonance and is illustrated in Figure 8.4(c) of \cite{Murray99}. Thus the differing gap shapes for interior versus exterior resonances allow a great deal of information about a perturbing planet to be gleaned from images of a disk that displays gaps, even if the planet itself remains unseen.

% P2
Following the same analysis that is done for the interior resonance, we make a Gaussian fit to the gap that can be seen in the histogram for the region closest to the planet (see top panel of Figure \ref{Fig:innerEI0-hists}) to obtain the width and the mean location of the 2:1 MMR gap. We discuss the results of these measurements in Sections \ref{Sec:WvsMP} and \ref{Sec:XCvsMP}.

\begin{figure*}
    \centering
    \subfigure[Same as Figure \ref{Fig:innerEI0-disk} except that the planet is given some eccentricity ($e\sim0.05$). Similar to the case of the interior resonance with the perturbing planet on a non-circular orbit (Figure \ref{Fig:outerERI0-disk}), an extra gap appears in the disk with location corresponding to the 3:1 MMR with a planet interior to the disk.]{%
    \includegraphics[totalheight=0.25\textheight]{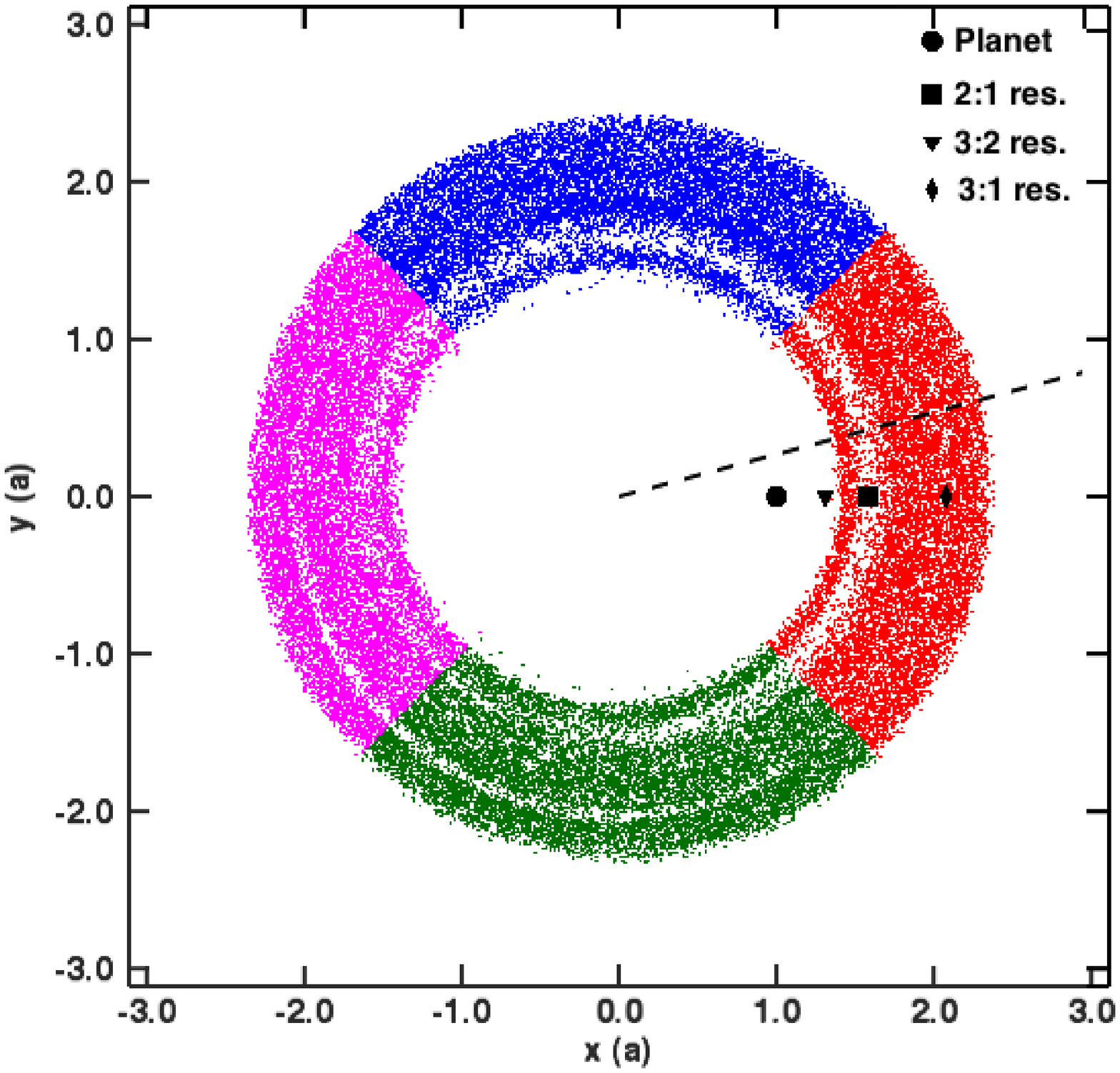}
    \label{Fig:innerERI0-disk}}
\hfill
    \subfigure[Histograms corresponding to the different segments of the disk shown in Figure \ref{Fig:innerERI0-disk}.]{%
    \includegraphics[totalheight=0.46\textheight]{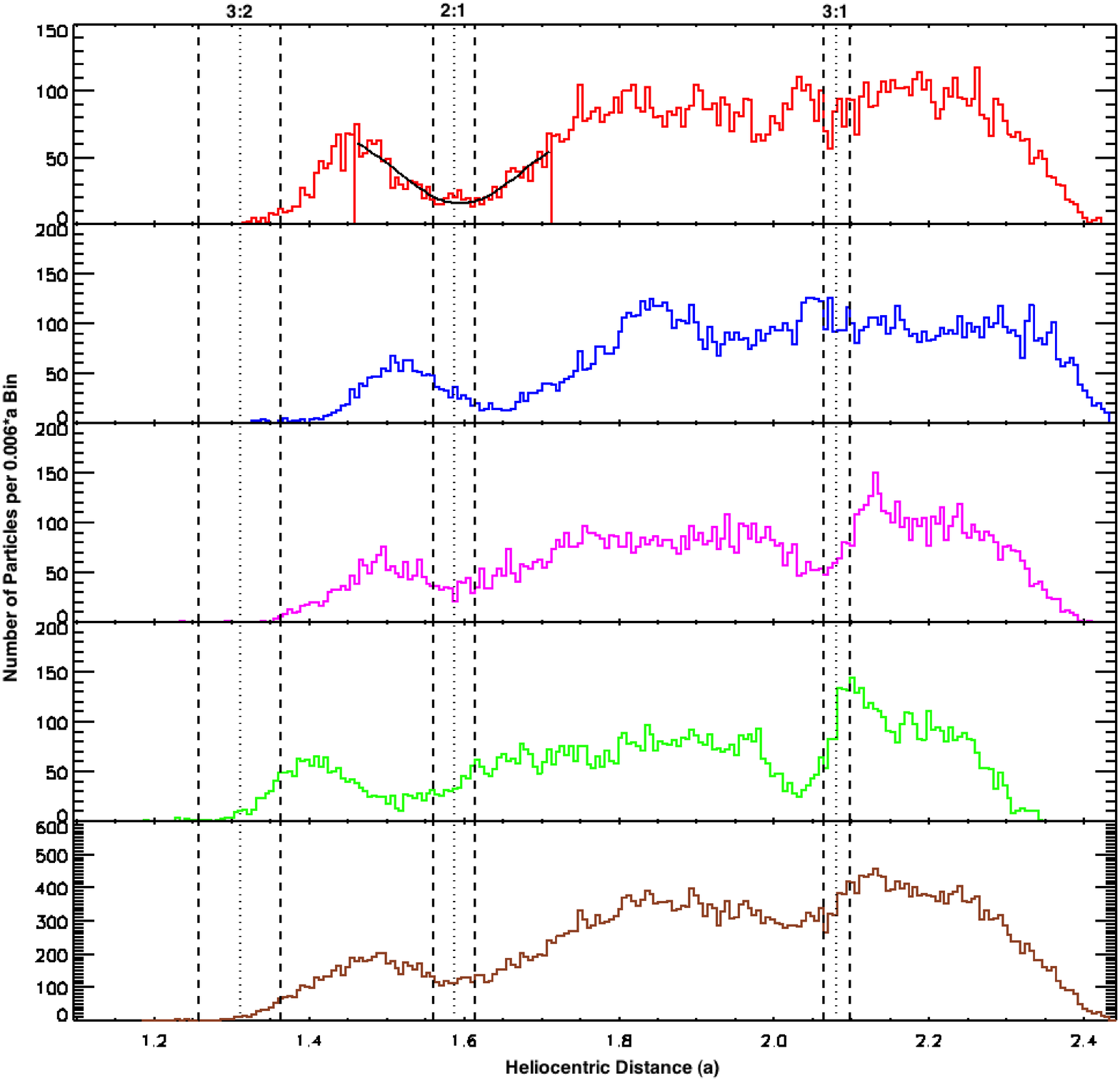}
    \label{Fig:innerERI0-hists}}
\caption{Same as Figure \ref{Fig:innerEI0-disk-hists} but with planet's eccentricity increased to $e = e_J = 0.0489$.}
\label{Fig:innerERI0-disk-hists}
\end{figure*}

% P3
Figure \ref{Fig:innerEI0-disk-hists} is obtained when the perturbing body (the planet) has no eccentricity. When the planet's orbital eccentricity is increased to that of Jupiter, the arc in the 2:1 gap remains easily visible but extends further, becoming more annular in shape. This is shown by Figure \ref{Fig:innerERI0-disk-hists}. In addition, the 3:1 feature observed in the case of the interior resonance (Figure \ref{Fig:outerERI0-disk-hists}) also becomes more prominent, and is explained in Section \ref{Sec:Lindblad}.

\section{Discussion}
\label{Sec:Discussions}

The main purpose of this study is to provide the means of determining the properties of a perturbing planet in the case where an MMR gap is observed in a debris disk. This includes both the characterization of detected planets, as well as providing information on the mass and location of planets which remain as-yet unseen. We consider now how the results of our simulations provide insight into this issue.

\subsection{Asymmetries in MMR Gaps}
\label{Sec:Asymm}

Our simulations suggest that the shape of an MMR gap is different for interior and exterior resonances. Therefore, by looking at the shape of a gap in a planetesimal disk we are not only able to immediately determine which side of the disk the perturbing planet is, the azimuthal asymmetry of the MMR gaps also allows us to easily distinguish between gaps formed due to MMRs with a planet that lies outside the gaps and those formed by planets in the gaps for which the gaps are azimuthally symmetric (such as the ones seen in the HL Tau disk). The only exception, according to our simulations, is the case of exterior resonance with a planet having non-zero orbital eccentricity for which we obtain nearly annular gaps at the 2:1 resonances. We note here that it is possible that increasing the eccentricity of the planet would result in the gap to eventually become completely annular, although we do not see that in a single sample simulation with $e=0.1$. However, even in this case, we still observe an asymmetric feature at the 3:1 MMR with the planet. The presence of this additional gap at the 3:1 resonance, which occurs for both interior and exterior resonances, is indicative of the planet having non-zero eccentricity. We leave examining the effect of planet eccentricity on debris disk structure to a future paper.

\subsection{Minimum and Maximum Detectable Planet Masses}
\label{Sec:MinMaxMP}

\begin{figure*}
    \centering
    \subfigure[No MMR gap is obvious in this simulated disk due to the small mass of the perturbing planet ($M=1.0 ~M_\oplus$)]{%
    \includegraphics[totalheight=0.25\textheight]{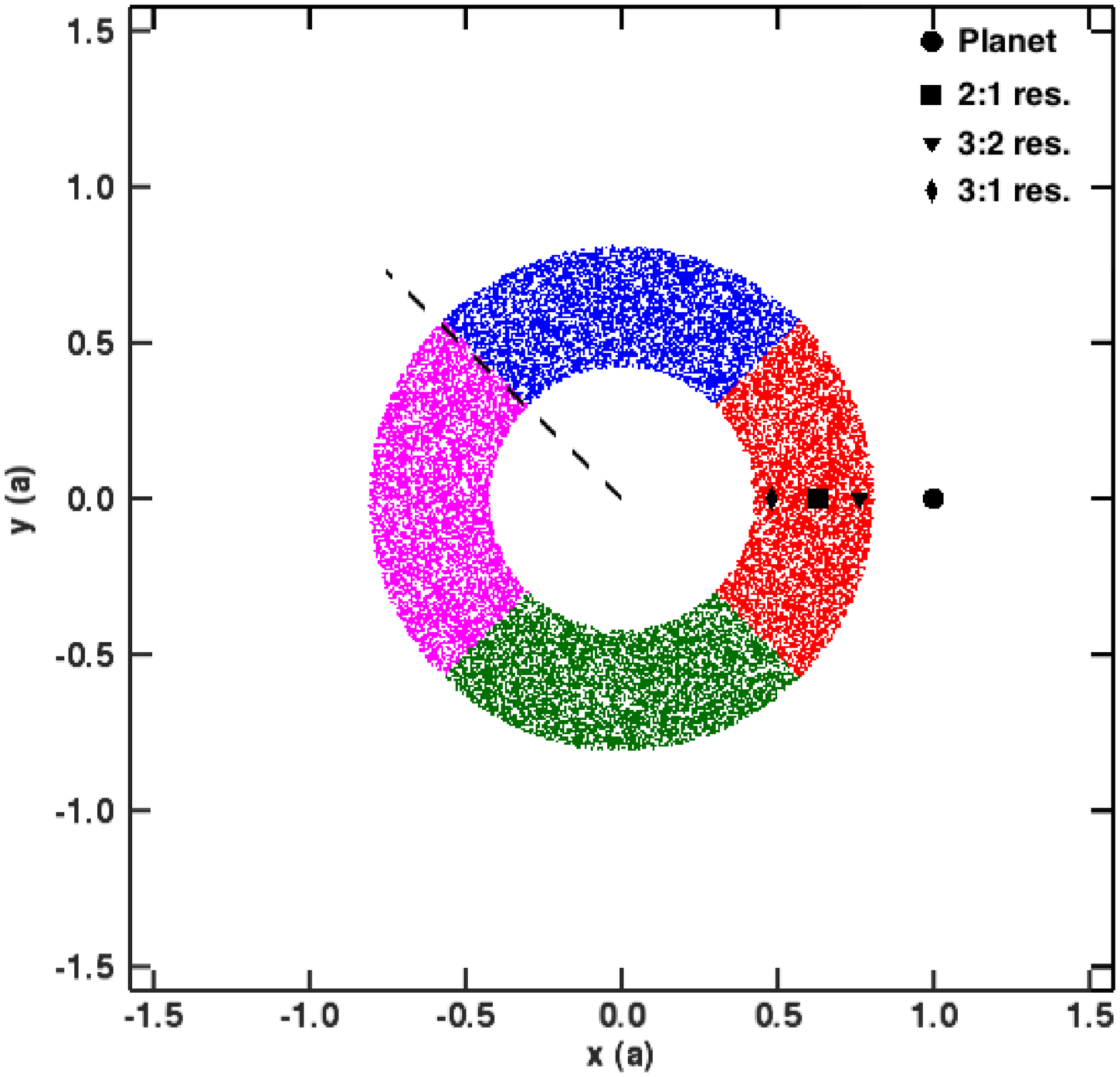}
    \label{Fig:MinM-disk}}
\hfill
    \subfigure[Histograms for the four segments are noisy, but putting them together averages the noise out and reveals two narrow gaps at the 2:1 and 3:2 MMR with the planet.]{%
    \includegraphics[totalheight=0.46\textheight]{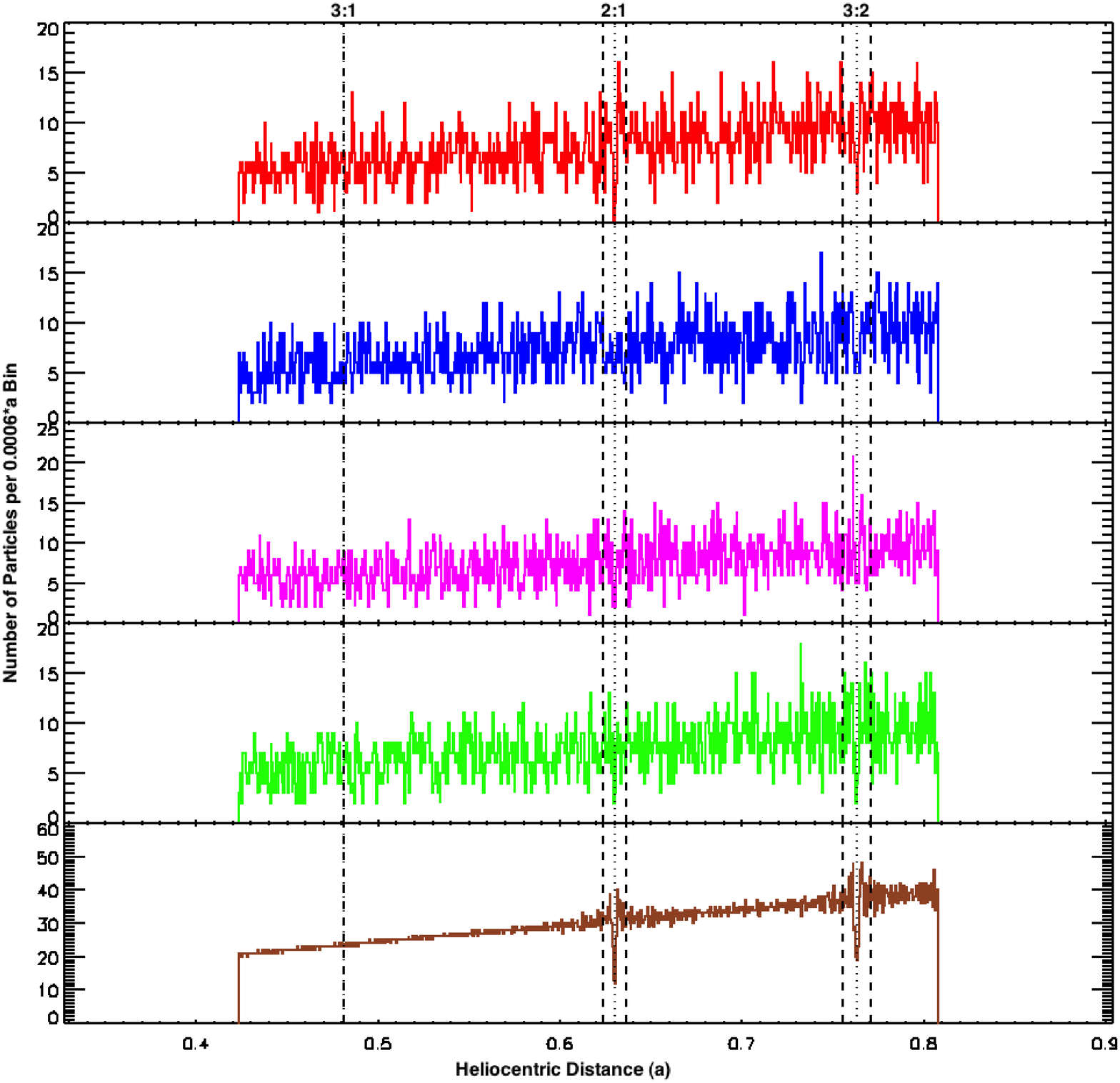}
    \label{Fig:MinM-hists}}
\caption{Though MMR gaps are not obvious in this example with a $1.0 ~M_\oplus$ perturbing planet, examining the last histogram reveals two dips at the 2:1 and 3:2 MMR with the planet. Such small gaps might not be detectable in a telescopic image of the disk; nevertheless our simulations show that even an Earth-mass planet is able to open gaps in disks through MMRs.}
\label{Fig:MinM-disk-hists}
\end{figure*}

We find that planets as small as $M=1.0 ~M_\oplus$ can produce MMR gaps (see Figure \ref{Fig:MinM-disk-hists}), but the practical lower limit on the mass of the planet that can open a detectable gap depends on the resolving capabilities of the observational facility taking the images. We are not aware of any disk with features meeting the above descriptions (i.e., disks with observed azimuthally asymmetric MMR gaps) that have yet been reported, but given the resolutions obtained by current facilities, we expect such features to be discovered in the near-future. For illustration, the disk around HL Tau has a radius of 80 AU \citep{Kwon11} and the recent ALMA image of the disk shows gap features as small as 5 AU across \citep{Tamayo15}. For comparison, a 3 Jupiter-mass planet on a circular orbit just outside the edge of a planetesimal disk of similar size as the HL Tau disk would create a 2:1 resonance gap that is $W = (0.013 \times 3.0 + 0.020) \times 80.0 \sim 4.7$ AU wide, comparable to the observed gaps that have been associated with planet formation, and easily distinguishable. It must be noted that ALMA probes particles that are $\sim$ 1 mm and smaller in size, but here we model much larger bodies. Nonetheless, millimeter-size particles are nearly unaffected by the PR drag and thus we obtain similar results when we consider disks that are entirely composed of $1$ mm dust and when collisional dust production is ignored. Yet future work is needed to study MMR gaps when collision among millimeter and sub-millimeter particles and radiative forces are taken into consideration. On the other hand, if the planet becomes very massive, the edge of the disk thins under its perturbations as larger mass planets excite the particles and move them to higher eccentricity orbits. In our simulations, the disk becomes too heavily eroded to distinguish the 2:1 MMR at $6-7 ~M_J$, making this the practical upper limit for the use of this technique for characterizing extrasolar disk-planet systems.

\subsection{Libration Width vs. Planet Mass}
\label{Sec:WvsMP}

\begin{figure*}
    \centering
    \includegraphics[totalheight=0.3\textheight]{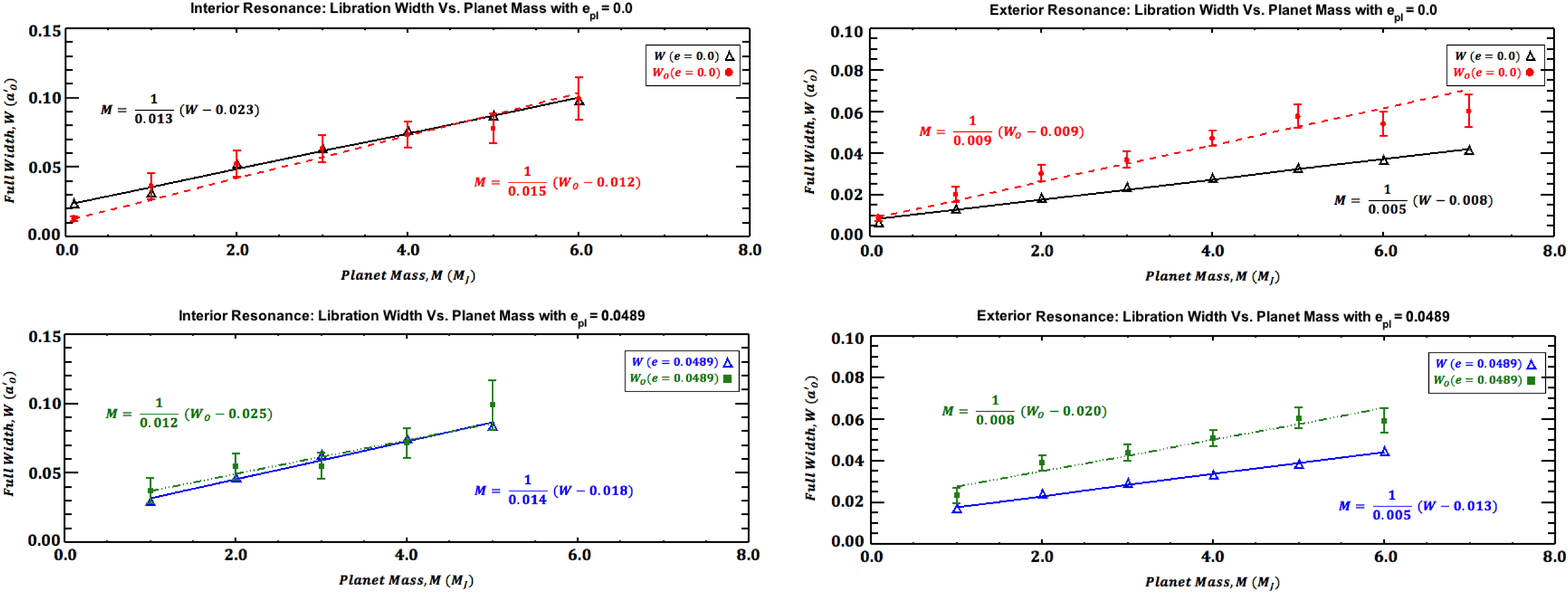}
    \caption{Libration width versus planet mass. Left panels: interior resonance with $e=0.0$ (top) and $e=0.0489$ (bottom). Right panels: exterior resonance with $e=0.0$ (top) and $e=0.0489$ (bottom). $W$ and $W_o$ are the theoretical and the simulated width of the gap, respectively, and are given in units of the observed mid-location of the gap ($a^\prime_o$). The vertical axis is the full width of the gap, $2 \sigma$ of the fitted Gaussian for the simulations.}
    \label{Fig:WvsMP}
\end{figure*}

% P1
More massive planets are expected to open wider MMR gaps in disks. In Figure \ref{Fig:WvsMP} we illustrate this effect by plotting libration width versus planet mass for interior (left) and exterior (right) resonances with planet's eccentricity $e=0.0$, shown by the top two panels, while the bottom two panels show our results when the planet's eccentricity is increased to $e=0.0489$. The different colors and symbols used and their least-square linear fits in each panel represent results obtained analytically using Equation \ref{Eq:delta_ap} and from the simulations. It is clear that particularly in the case of interior resonances (left two panels), there is nearly perfect agreement between the 2:1 MMR gap widths obtained from the simulations and the ones calculated analytically while there is an offset between the measured and analytic results in the case of exterior resonances (right panels). Furthermore, as expected, increasing the planet's mass also increases the width of the MMR gap in all cases. Thus the mass of the perturbing planet, seen or unseen, can be estimated from the width of the resonant gap it produces.

% P2
Equations \ref{Eq:outerEI0WvsMP} through \ref{Eq:innerERI0WvsMP} are obtained from least square fits to the values we get from our simulations (see Figure \ref{Fig:WvsMP}). These equations allow one to easily calculate the planet's mass if the 2:1 MMR gap width can be measured observationally. To simplify the calculation, we normalize our results again, this time making the observed mid-location of the gap equal to one unit of distance. We do this so that the equations that are provided from this point forward can also be applied to cases where the planet is unseen but an MMR gap is detected in the debris disk. \\

(A) Interior Resonance with $e=0.0$:

\begin{equation}
\label{Eq:outerEI0WvsMP}
M = \frac{1}{0.015}(W_o-0.012),
\end{equation}

(B) Interior Resonance with $e=0.0489$:

\begin{equation}
\label{Eq:outerERI0WvsMP}
M = \frac{1}{0.012}(W_o-0.025),
\end{equation}

(C) Exterior Resonance with $e=0.0$:

\begin{equation}
\label{Eq:innerEI0WvsMP}
M = \frac{1}{0.009}(W_o-0.009),
\end{equation}

(D) Exterior Resonance with $e=0.0489$:

\begin{equation}
\label{Eq:innerERI0WvsMP}
M = \frac{1}{0.008}(W_o-0.020),
\end{equation}

\noindent where $M$ is the planet mass (in Jupiter masses, $M_J$) and $W_o$ is the observed width of the gap (in units of the distance between the star and the observed gap).

% P3
The error bars are calculated by taking three main sources of uncertainty into account that are added in quadrature:

% P4
1) The Gaussian fit to the histogram is made by least-square fitting using Interactive Data Language (IDL) and the goodness of fit is recorded as one source of uncertainty. This value is generally small in the examples we tried.

% P5
2) Since the Gaussian fit is made to points that mark half the bin size, the measurements have uncertainties that are affected by the choice of the bin size (taken to be 0.006 times the planet's distance from the star in this work). However, the uncertainty in each bin position goes as $\sqrt{n}$, where $n$ is the number of particles in the bin; and since we chose the bin size such that on average each bin contains not more than $0.25\%$ of the total number of particles in the disk, the uncertainty due to the finite size of the bin is usually not significant.  

% P6
3) Examining the shapes of the MMR gaps in our simulations revealed that gaps are not always perfectly Gaussian in shape. This is illustrated, for instance, in the top panel of Figure \ref{Fig:outerEI0-hists} where the gap is higher at one end. Therefore, in order to find the width of the gap, the Gaussian fit is made three times by normalizing to either side and also without normalization. The standard deviation between the three values obtained is then taken as the uncertainty in the gap width and is the dominant source of uncertainty in our calculations. It must be noted that our results for the gap widths are shown in terms of each gap's standard deviation while an observer measuring the width of the gap might define the edges by measuring the peak brightness in the disk near the edge, then locating the radius in the gap at which the brightness of the disk is half that value. This method of using the "half-maximum radius" is used by \cite{Chiang09} for simulations of the Fomalhaut disk. Nevertheless, the two quantities are related by a simple formula shown by Equation \ref{Eq:FWHM}:

\begin{equation}
\label{Eq:FWHM}
FWHM = 2\sqrt{2 \times \ln{2}} ~ \sigma,
\end{equation}
where $FWHM$ is the full width at half maximum and $\sigma$ is the standard deviation ($W_o$ in our equations).

\subsection{Resonant Location Versus Planet Mass}
\label{Sec:XCvsMP}

\begin{figure*}
    \centering
    \includegraphics[width=0.85\textwidth, height=0.55\textwidth]{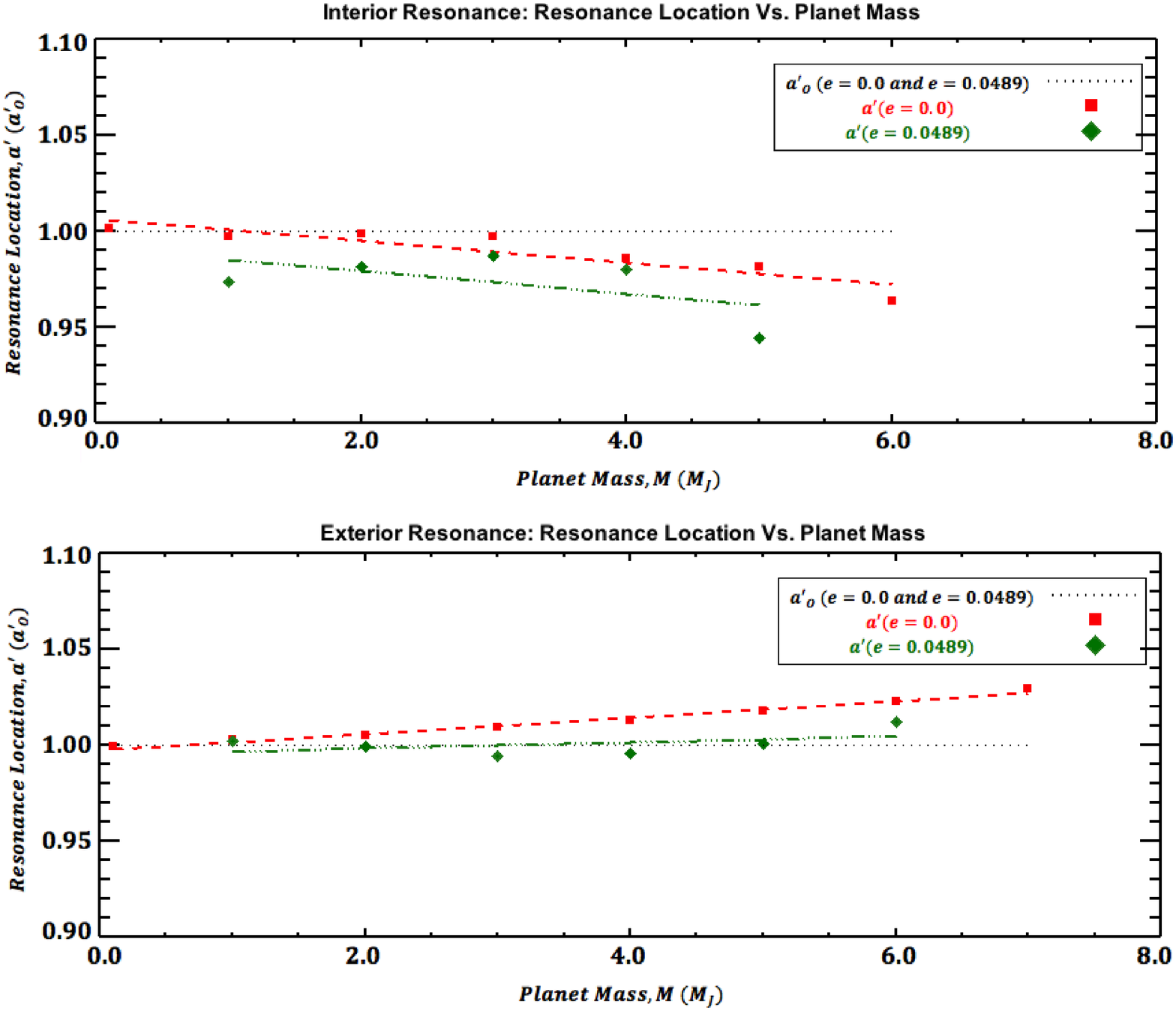}
    \caption{The effect of planet's mass and eccentricity on the resonant location for interior (top) and exterior (bottom) resonances. The red squares and the green diamonds show the theoretical gap locations for cases with $e=0.0$ and $e=0.0489$, respectively while the dotted lines show the locations of the observed gaps.}
    \label{Fig:XCvsMP}
\end{figure*}

% P1
Examining the 2:1 MMR gaps in our histograms suggest that there is a shift in the position of the gap (the peak of the Gaussian fit) from the nominal resonance location (the dotted lines on the histograms). In fact, our simulations show that the offset from the gap's theoretical position is proportional to the mass of the planet. This is shown in Figure \ref{Fig:XCvsMP} where the top and bottom panels are for interior and exterior resonances, respectively. The red squares show the location of the 2:1 resonance with a zero-eccentricity planet while the diamond symbols in green are for the case of higher eccentricity planet with $e=0.0489$ and the dotted lines denote the location of the nominal resonance calculated using Equation \ref{Eq:MMR_Def_a} for interior and exterior resonances. The uncertainties in the observed gap locations are calculated in the same manner as those for the gap widths, explained in Section \ref{Sec:WvsMP}. 

% P2
In addition to a shift in the resonance location, we note from Figure \ref{Fig:XCvsMP} that the MMR gaps always tend to shift toward the planet as can be seen from the negative slope in the top panel and the positive slope in the bottom panel. We find that there is small contribution from the planet's eccentricity in shifting the locations of the MMR gaps. In addition, our results indicate that the location of the 2:1 gap agrees better with the theoretical prediction for the higher eccentricity planet in the exterior resonance case.

\subsection{Finding the Planet's Semimajor Axis based on the Observed Gap Width and Location}
\label{Sec:a_pl}

% P1
If the mid-location and the width of an MMR gap can be obtained through observations, this information can be used to calculate the planet's semimajor axis. This could be useful for calculating the orbit of a known planet or determining the location of an unseen one. This calculation is done by first measuring the distance from the star to the center of the gap, $a_o^\prime$ and then finding the gap width in units of the observed star-gap separation. Then Equations \ref{Eq:outerEI0WvsMP} through \ref{Eq:innerERI0WvsMP} can be used to obtain the planet's mass. The theoretical location of the gap, $a^\prime$, (i.e. the nominal resonant location) can then be found through the following equations:

(A) Interior Resonance with $e=0.0$:

\begin{equation}
\label{Eq:outerEI0Shift}
a^\prime = -0.003~M+0.002+a_o^\prime,
\end{equation}

(B) Interior Resonance with $e=0.0489$:

\begin{equation}
\label{Eq:outerERI0Shift}
a^\prime = -0.002~M-0.015+a_o^\prime,
\end{equation}

(C) Exterior Resonance with $e=0.0$:

\begin{equation}
\label{Eq:innerEI0Shift}
a^\prime = 0.004~M-0.001+a_o^\prime,
\end{equation}

\begin{figure*}
    \centering
    \includegraphics[width=0.85\textwidth, height=0.55\textwidth]{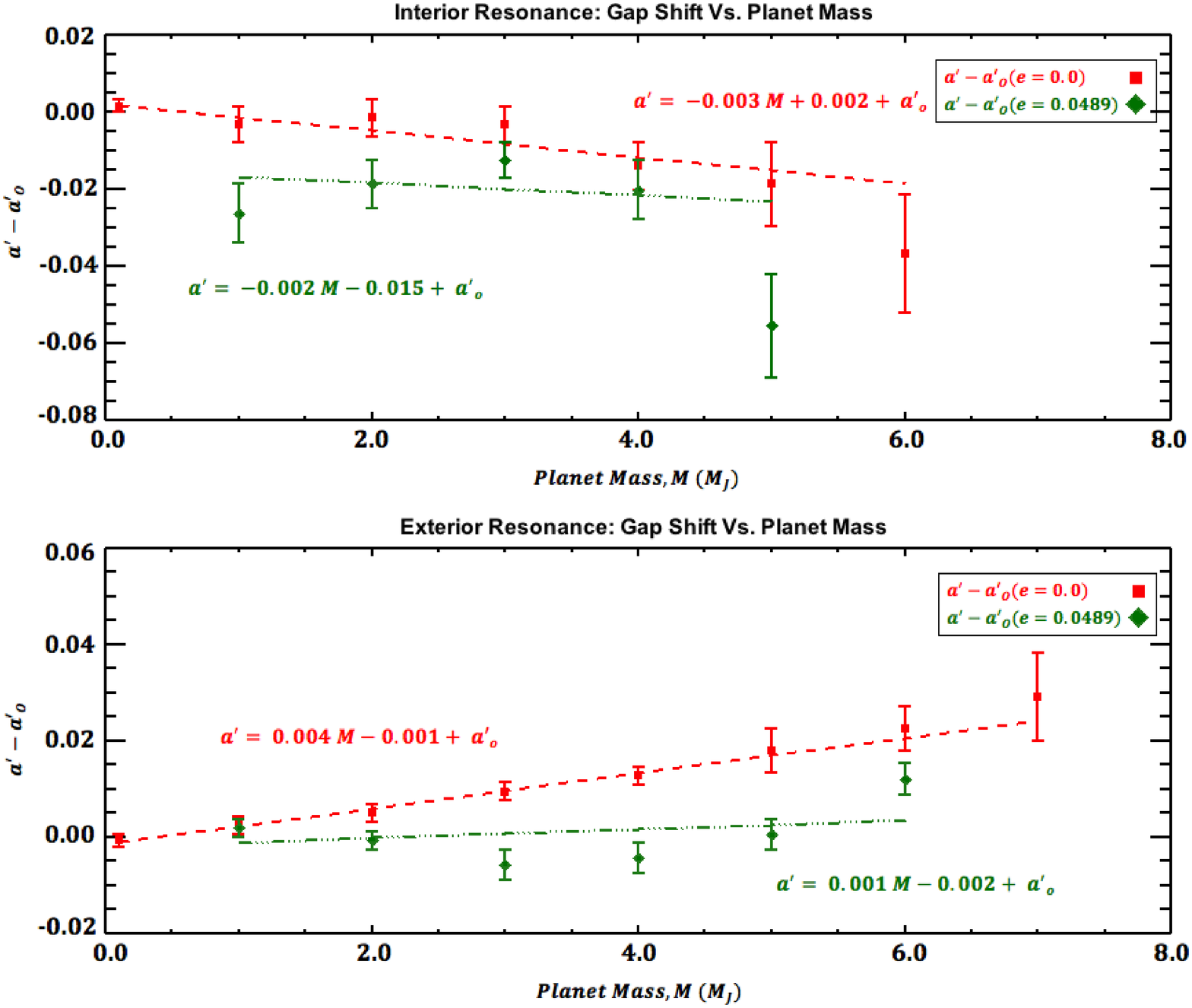}
    \caption{The shift in the gap location versus the planet's mass for interior (top) and exterior (bottom) resonances for two different planet eccentricities. The symbols used are the same as those in Figure \ref{Fig:XCvsMP}. $a^\prime$ and $a_o^\prime$ are the theoretical and observed location of the gap, respectively.}
    \label{Fig:ShiftvsMP}
\end{figure*}

(D) Exterior Resonance with $e=0.0489$:

\begin{equation}
\label{Eq:innerERI0Shift}
a^\prime = 0.001~M-0.002+a_o^\prime,
\end{equation}

\noindent where $M$ and $a^\prime$ have units of Jupiter mass ($M_J$) and $a_o^\prime$, respectively. Finally, the semimajor axis of the planet can be calculated using Equation \ref{Eq:MMR_Def_a}. 
Equations \ref{Eq:outerEI0Shift} through \ref{Eq:innerERI0Shift} were obtained by subtracting the observed gap location from its theoretical location in our simulations. This is shown by Figure \ref{Fig:ShiftvsMP} which can be used to obtain the theoretical MMR gap location from which the planet's semimajor axis is found.

\subsection{The Effect of Planet's Orbital Eccentricity and Lindblad Resonances}
\label{Sec:Lindblad}

When the planet is given a small but non-zero eccentricity, the 2:1 MMR properties remain largely unaffected. However, tightly wound spiral waves originating from the 3:1 MMR appear in some cases. These features are potentially valuable sources of information about the disk's properties, but are far more challenging to detect in real telescopic images than the gaps associated with the MMRs that we have been discussing so far. In fact, they often appear at scales below those shown in the previous figures, their primary visible tracer being a narrow arc-like gap at the 3:1 interior or exterior resonance.

\begin{figure*}
    \centering
    \includegraphics[totalheight=0.45\textheight]{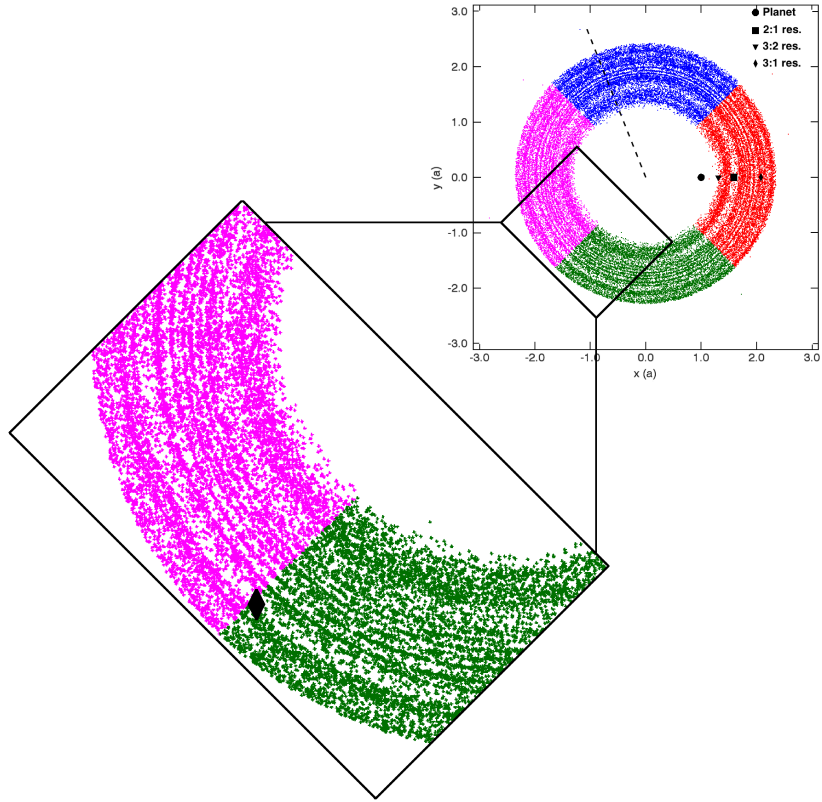}
    \caption{Spiral patterns appear commonly in our simulations when the perturbing planet has non-zero orbital eccentricity. These are likely forced eccentricity waves originating at Lindblad resonances. Our simulations indicate that these waves are generated at a location corresponding to the the 3:1 MMR with the planet, marked by a diamond on this figure.}
    \label{Fig:Spiral}
\end{figure*}

We interpret these features as forced eccentricity waves originating at Lindblad resonances \citep{Shu84}. These waves are similar to those seen in Saturn's rings \citep[e.g.][]{Holberg82, Lane82} and have the characteristic decrease in wavelength as one moves away from the resonance \citep{Murray99}. This is illustrated by Figure \ref{Fig:Spiral} which shows a series of density waves originating from the 3:1 resonance.

These waves are seen to be produced primarily at the 3:1 MMR for both internal and external planetary perturbers and are likely the reason we see an extra feature at the 3:1 resonance with the planet whenever the planet is given non-zero eccentricity in our simulations (see Figures \ref{Fig:outerERI0-disk-hists} and \ref{Fig:innerERI0-disk-hists}). These coincide with the location of the $m=2, ~k=\pm1, ~p=0$ inner/outer Lindblad resonances, which are associated with similar waves in Saturn's rings. Though such structures are a rich source of information about the planet and the disk itself, the propagation of waves in real systems depends on effects such as self-gravity and collisions \citep{Fridman99} which are not modeled here, and we will leave their examination for future work.

Although the 3:1 MMR gaps are narrower and more difficult to measure than the 2:1 gaps, if they can be detected, their relative location with respect to that of the 2:1 resonance can be used to also distinguish interior from exterior resonances as the 3:1 resonance gap is formed farther from the planet than the 2:1. We will leave detailed examination of resonant interaction between disk particles and higher eccentricity planets to a follow-up paper.

\subsection{Disk Optical Depth and Gap Contrast}
\label{Sec:Contrast}

\begin{figure*}
    \centering
    \includegraphics[width=0.85\textwidth, height=0.55\textwidth]{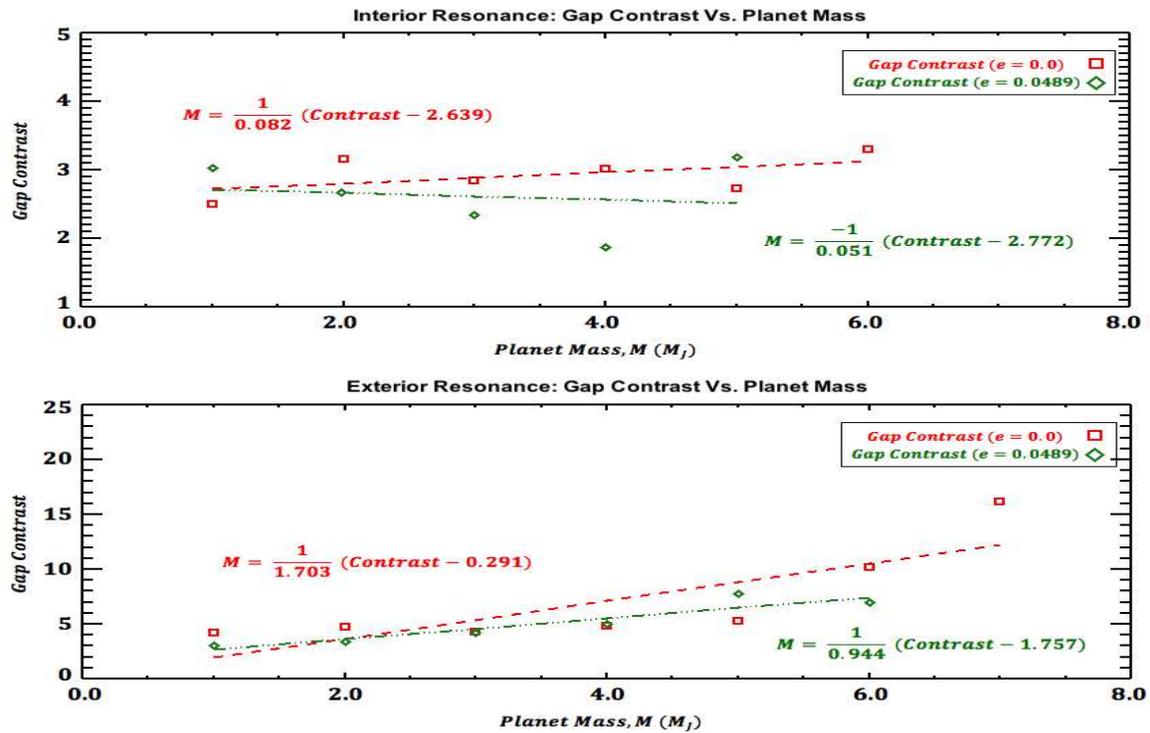}
    \caption{MMR gap contrast vs. planet's mass (in Jupiter mass) for interior (top) and exterior (bottom) resonances.}
    \label{Fig:ContrastvsMP}
\end{figure*}

Though the gap widths can easily be measured in simulations, this may prove more difficult observationally. In particular, the optical depth in the gap versus the disk as a whole will determine the amount of contrast in the image. Figure \ref{Fig:ContrastvsMP} shows the ratio of the average disk surface density to that at the deepest part of the 2:1 MMR gap (the "contrast"), as a function of planet mass. Note that the contrast we refer to here is the contrast in planetesimal surface density, not dust surface density, because we are considering dust-poor systems. Because of the difficulty of defining the edges of the gaps, we do not try to define an edge-to-center contrast. Though higher mass planets form larger gaps, the contrast is somewhat diminished by particles "bleeding" in from the edges of the gaps. There is a trend for the contrast to increase with mass for a planet that is interior to the disk (Figure \ref{Fig:ContrastvsMP}, bottom panel) but there is little effect for an external perturber (Figure \ref{Fig:ContrastvsMP}, top panel). Thus the gap contrast itself can provide a measure of the planet mass in some cases, though this approach is observationally more difficult. Nonetheless, the contrast remains at large values through the range of planet masses considered here, indicating that the MMR features discussed here should be detectable if the disks themselves are.

We note from our results that the contrast grows almost linearly with planet mass, except for exterior resonance with a planet on a circular orbit in which case there seems to be an exponential trend. However, we feel it is unwarranted to fit an exponential to this case for two reasons: First, the appearance of exponential growth is only present for one of the four cases (exterior resonance with a planet having $e=0$) and rests on only the two rightmost points. Moreover, those two points are near the largest masses beyond which the gaps disappear due to heavy erosion of the disk edge; and so the graphs should not be extrapolated beyond the presented maximum value.

\begin{figure*}
    \centering
    \includegraphics[totalheight=0.45\textheight]{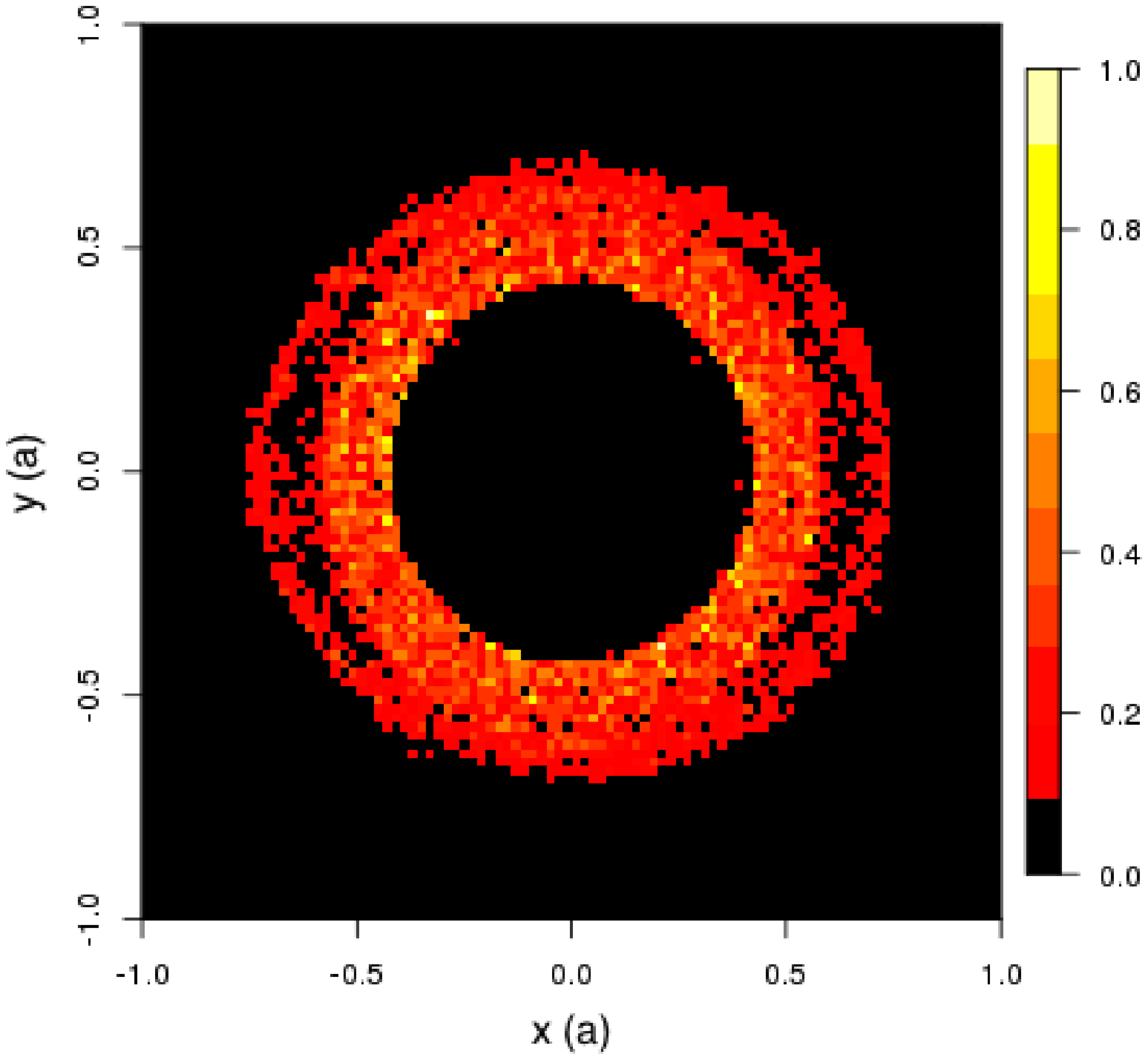}
    \caption{Same as Figure \ref{Fig:outerEI0-disk} for interior resonance except that particles in $x$ and $y$ are binned and assigned a color based on the total emission from each bin, normalized to the peak intensity. Darker colors correspond to less emission.}
    \label{Fig:obs-outer}
\end{figure*}

\begin{figure*}
    \centering
    \includegraphics[totalheight=0.45\textheight]{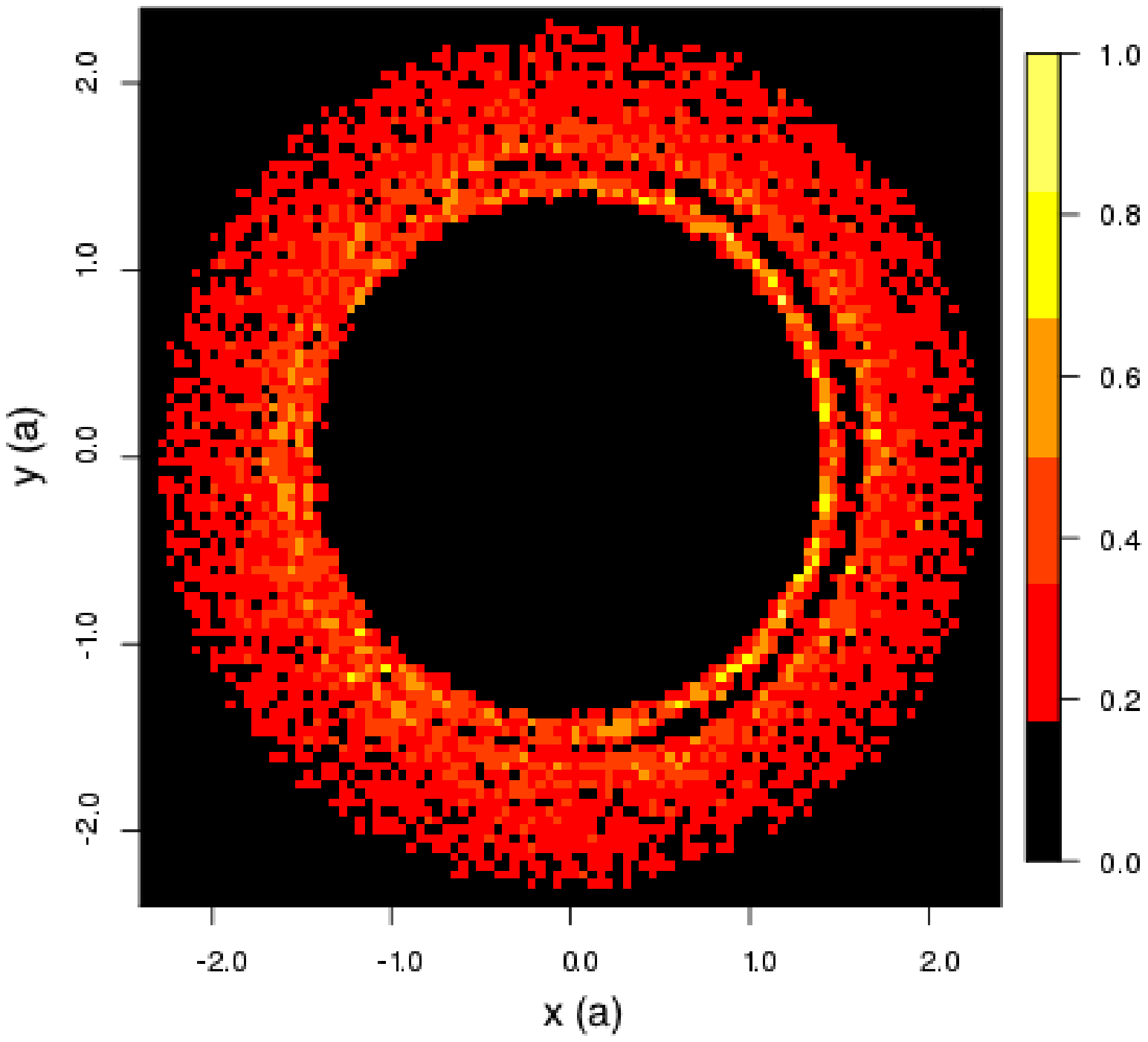}
    \caption{Same as Figure \ref{Fig:obs-outer} but for the case of exterior resonance. The same disk is shown in Figure \ref{Fig:innerEI0-disk}.}
    \label{Fig:obs-inner}
\end{figure*}

Simulated observations of the disks shown earlier in Figures \ref{Fig:outerEI0-disk} and \ref{Fig:innerEI0-disk}, are illustrated in Figures \ref{Fig:obs-outer} and \ref{Fig:obs-inner} assuming an inverse-square dependence of particle emission on distance from the central star. It must be noted again that in this study, we assume that disks are optically thin and are largely free of collisionally produced or other sources of dust. These simple figures are free of additional noise that would certainly be present in real observations. The pixel size is chosen to match ALMA's highest resolution at 1~mm wavelengths (15 mas for the 16~km configuration) at the distance of HL~Tau (140 pc) if the disk's outer radius is 100~AU. Since the MMR gaps can still be seen in these images, we conclude that the MMR features described here are nominally within the reach of current observational facilities.

\section{Summary and Conclusions}
\label{Sec:SummConc}

% P1
We investigated the dynamical effects of a planet on a planetesimal disk through MMRs for both interior and exterior resonances. Our purpose is to use the observed properties of MMR structures to characterize the planet producing them, even if that planet remains as-yet undetected.

% P2
Structures arising from MMRs can be highly diagnostic of the properties of the planet disturbing the disk particles. MMR gaps become wider as the planet's mass increases; there is a linear relationship between the gap width and the perturbing planet's mass that agreed well with theoretical calculations. Therefore, measurement of the width of a 2:1 MMR gap would help determine the perturbing planet's mass, even if it remains unseen. We find gaps at the 2:1 and 3:2 MMRs even for a planet as small as $1.0 ~M_\oplus$, although their small widths make them observationally more challenging to detect than those at Jupiter masses. On the other hand, at planet masses beyond $6-7 ~M_J$, the resonance structures are destroyed as the disk is eroded by the planet's growing Hill Sphere.

% P3
We found an offset in the gap's position in the disk from the nominal resonant location with more massive planets causing a larger shift in the observed location of the gap. Thus if the planet's location is already known, the shift from the theoretical location of the 2:1 MMR gap can be used to confirm the planet's mass, which can alternatively be calculated using the gap width. On the other hand, for cases in which the planet remains undetected, we proposed a set of equations that take the planet's mass, calculated using the gap width, to determine the planet's location based on that of the observed 2:1 gap.

% P4
We further extended our studies to simulations of systems in which the perturbing planet has non-zero orbital eccentricity. In this case, disk structures due to MMRs become more complicated and the 2:1 MMR gaps formed by an internal perturber become more annular in shape. Moreover, an extra arc-like feature was seen at the 3:1 MMR with the planet when the planet's eccentricity was increased and is associated with spiral waves generated at Lindblad resonances. Thus the appearance of an arc-shaped gap at the 3:1 resonance with the planet is indicative of the planet having non-zero eccentricity. If detected, it could also be used to distinguish interior from exterior resonance. This would be helpful, particularly for the case of exterior resonance with a non-zero eccentricity planet for which we found the 2:1 gaps to be azimuthally symmetric. We obtained nearly perfect agreement between the 2:1 gap widths measured analytically and through our simulations for a planet exterior to the disk with 0 or 0.0489 orbital eccentricity. On the other hand, the difference between the two measurements seems to grow with mass for a planet on a circular orbit interior to the disk while a systematic shift is seen when the planet's eccentricity is increased to that of Jupiter's in this case. All the simulations we report on here were performed for flat disks with the perturbing planet in the same orbital plane, but we found no significant difference in our results when the planet, and thus the particles, were given a small orbital inclination.

% P5
The results of our simulations indicate that the shapes of the gaps opened by the 2:1 MMRs are different for interior versus exterior resonances, with the former making two sectors at the planet's (inferior) conjunction and opposition while the latter forms a single arc at (superior) conjunction. Since direct detection of extrasolar planets still remains observationally challenging, detection of such structures in a planetesimal disk allows one to not only infer the presence of an unseen planet, the two distinct gap shapes would also make it possible to easily determine the relative location of the planet with respect to the disk and to distinguish MMR gaps from azimuthally symmetric gaps formed by a planet that is embedded in the disk. As ALMA and other facilities continue to advance the frontiers of extrasolar planetary science, the ability to detect and characterize unseen planets based on their effects on a more-easily observable disk will become an increasingly powerful tool.

\acknowledgments

The authors wish to thank the anonymous referee for valuable comments. This work was supported in part by the Natural Sciences and Engineering Research Council of Canada (NSERC).

\bibliography{main}

\end{document}